\documentclass[conference, letterpaper]{IEEEtran}
\IEEEoverridecommandlockouts
\usepackage{cuted}
\usepackage{cite}
\usepackage{bm}
\usepackage{comment}
\usepackage{amsmath,amssymb,amsfonts}
\usepackage{algorithm}
\usepackage{lettrine}
\usepackage{graphicx}
\usepackage{multirow}
\usepackage{algpseudocode}
\usepackage{diagbox}
\usepackage{graphicx}
\usepackage{subcaption}
\usepackage[letterpaper, left=0.625in, right=0.625in, bottom=1.02in, top=0.70in]{geometry}
\usepackage{slashbox}
\usepackage{textcomp}
\usepackage{xcolor}

\newcommand{\be}{\begin{equation}}
\newcommand{\ee}{\end{equation}}

\def\BibTeX{{\rm B\kern-.05em{\sc i\kern-.025em b}\kern-.08em
    T\kern-.1667em\lower.7ex\hbox{E}\kern-.125emX}}

\usepackage{scalerel}
\usepackage{tikz}
\usetikzlibrary{svg.path}
\definecolor{orcidlogocol}{HTML}{A6CE39}
\tikzset{
  orcidlogo/.pic={
    \fill[orcidlogocol] svg{M256,128c0,70.7-57.3,128-128,128C57.3,256,0,198.7,0,128C0,57.3,57.3,0,128,0C198.7,0,256,57.3,256,128z};
    \fill[white] svg{M86.3,186.2H70.9V79.1h15.4v48.4V186.2z}
                 svg{M108.9,79.1h41.6c39.6,0,57,28.3,57,53.6c0,27.5-21.5,53.6-56.8,53.6h-41.8V79.1z M124.3,172.4h24.5c34.9,0,42.9-26.5,42.9-39.7c0-21.5-13.7-39.7-43.7-39.7h-23.7V172.4z}
                 svg{M88.7,56.8c0,5.5-4.5,10.1-10.1,10.1c-5.6,0-10.1-4.6-10.1-10.1c0-5.6,4.5-10.1,10.1-10.1C84.2,46.7,88.7,51.3,88.7,56.8z};
  }
}

\newcommand\orcidicon[1]{\href{https://orcid.org/#1}{\mbox{\scalerel*{
\begin{tikzpicture}[yscale=-1,transform shape]
\pic{orcidlogo};
\end{tikzpicture}
}{|}}}}
\usepackage{hyperref}

\begin{document}

\title{Model-based Deep Learning for QoS-Aware Rate-Splitting Multiple Access Wireless Systems}

\author{Hanwen Zhang, \emph{Student Member, IEEE}, 
Mingzhe Chen, \emph{Senior Member, IEEE}, Alireza Vahid, \\ \emph{Senior Member, IEEE}, Feng Ye, \emph{Senior Member, IEEE},  and Haijian Sun, \emph{Senior Member, IEEE}
\thanks{
H. Zhang and H. Sun (hanwen.zhang@uga.edu, hsun@uga.edu) are with the School of Electrical and Computer Engineering, University of Georgia, Athens, GA, USA. 

M. Chen (mingzhe.chen@miami.edu) is with the Department of Electrical and Computer Engineering,
University of Miami, Coral Gables, FL, USA.  

A. Vahid (arveme@rit.edu) is with the Electrical and Microelectronic Engineering, Rochester Institute of Technology, Rochester, NY, USA.

F. Ye (feng.ye@wisc.edu) is with the Department of Electrical and Computer Engineering, University of Wisconsin-Madison, Wisconsin, WI, USA. 

This paper is an extended version of work that was presented in \cite{zhang2024model}.  This journal version includes the data rate QoS requirements for each user, providing a more detailed analysis on projected gradient descent and expanded results.

Dataset and codes of this paper will be available after paper acceptance. 
}}

\maketitle

\begin{abstract}

Next generation communications demand for better spectrum management, lower latency, and guaranteed quality-of-service (QoS). Recently, Artificial intelligence (AI) has been widely introduced to advance these aspects in next generation wireless systems. However, such AI applications suffer from limited training data, low robustness, and poor generalization capabilities. To address these issues, a model-driven deep unfolding (DU) algorithm is introduced in this paper to bridge the gap between traditional model-driven communication algorithms and data-driven deep learning. Focusing on the QoS-aware rate-splitting multiple access (RSMA) resource allocation problem in multi-user communications, a conventional fractional programming (FP) algorithm is first applied as a benchmark. The solution is then refined by the application of projection gradient descent (PGD). DU is employed to further speed up convergence procedure, hence improving the efficiency of PGD. Moreover, the feasibility of results is guaranteed by designing a low-complexity projection based on scale factors, plus adding violation control mechanisms into the loss function that minimizes error rates. Finally, we provide a detailed analysis of the computational complexity and analysis design of the proposed DU algorithm. Extensive simulations are conducted and the results demonstrate that the proposed DU algorithm can reach the optimal communication efficiency with a mere $0.024\%$ violation rate for 4 layers DU. The DU algorithm also exhibits robustness in out-of-distribution tests and can be effectively trained with as few as 50 samples. 
\end{abstract}


\section{Introduction}
% \lettrine[lines=2]{V}{ehicular} communication systems are recognized as pivotal components of intelligent wireless communication networks \cite{robey2020model}. However, the vast expansion of vehicle-to-everything (V2X) applications and services introduces substantial communication loads, thereby complicating resource management significantly. Three main challenges need to be addressed for the optimal resource allocation: high mobility, low-latency communication, and quality-of-service (QoS). High vehicle mobility requires effective algorithms that can keep up with dynamic channel characteristics. Concurrently, the resource management must be designed to minimize V2X communication latency. Furthermore, interference managment strategies are required to improve vehicular network QoS. Existing research has explored various approaches to mitigate the impact of interference on communications, including spectrum management \cite{zhang2024transformer}, dynamic spectrum access \cite{zhuang2018large}, advanced multiplexing techniques \cite{mathur2021survey}, and power control measures \cite{alizadeh2023power}. These strategies are integral to maintaining efficient and reliable communication in dense vehicular environments.

\lettrine[lines=2]{N}{ext} generation communication systems are bringing more potential applications and challenges. In particular, the vast expansion of edge devices and mobile users introduces substantial communication loads, thereby complicating underlying resource management significantly. Three main challenges need to be addressed for the optimal resource allocation: spectrum management, low-latency communication, and quality-of-service (QoS) guarantee.  Furthermore, interference management strategies are required to improve future communication network QoS. Existing research has explored various approaches to mitigate the impact of interference on communications, including spectrum management \cite{zhang2024transformer}, dynamic spectrum access \cite{zhuang2018large}, advanced multiplexing techniques \cite{mathur2021survey}, and adaptive power control \cite{alizadeh2023power}. Recently, rate-splitting multiple access (RSMA) has been studied as a potential and prominent strategy for interference management in communication systems \cite{zhang2024model,mao2018rate,yang2021optimization,clerckx2023primer,clerckx2016rate,clerckx2024multiple,xiao2023joint,dizdar2023rsma}. RSMA divides the data stream into two components: a common stream and a private stream. The common stream is designed to broadcast data to all users, while the private data stream is dedicated to each individual user \cite{zhang2024model,mao2018rate,yang2021optimization,clerckx2023primer,clerckx2016rate,clerckx2024multiple,xiao2023joint,dizdar2023rsma}. 
For example, in the context of vehicular communications, the RSMA common data stream typically includes map updates, traffic congestion alerts, and traffic light status, while the private data stream carries user-specific information such as route planning, car control, and social media messages \cite{zhang2024model}.  
The stream division has high spectrum efficiency as it applies partial spectrum resources for common data while enabling private communications with the remaining spectrum resources. 
% While the same time, the problems given in vehicular communication have better performance on QoS and mobile communication, the communication delay is still a challenge.
% Assuming delay is the major challenged tackled in this papaer, I'd suggest addding one paragraph here, introducing the proposed solution, i.e., the formulated problem. The machine learning part is a technical approach, which should be introduced after understanding the proposed problem.

To take full advantage of RSMA in spectrum efficiency,  a system requires an instant strategy for making resource allocation and interference management decisions. Leveraging the vast amount of data, deep learning methods have been introduced to enhance resource allocation strategy making in communication systems \cite{gong2023edge}. In particular, data-driven deep learning models, such as convolutional neural networks (CNN) \cite{zhou2022deep,wang2020deep}, long short-term memory (LSTM) \cite{zhou2022deep} and graph neural networks (GNN) \cite{zhang2022learning,li2023graph} are extensively utilized in communication researches. However, most of the existing data-driven deep learning models rely heavily on the distributions of the training data \cite{liu2021towards,robey2020model}, which can lead to unpredictable or unstable outputs given data of varying distributions. Moreover, the model weights, primarily shaped by implicit relationships and potentially prone to over-fitting, do not necessarily adhere to underlying physical principles, hence jeopardizing results from pure data-driven approaches. In comparison, model-based deep learning offers a promising solution to those limitations encountered in data-driven approaches \cite{shlezinger2023model,samuel2019learning,nguyen2020deep,zhang2023joint,xia2023deep,schynol2023coordinated,zhang2022deep}. Model-based deep learning usually constructs interpretable neural network structure with fewer trainable parameters, constrained within a well-defined framework based on physical reality. As a result, a model-based approach requires less data while achieving more robust outcomes. Model-based deep learning can manifest in several forms, including deep unfolding (DU), neural building blocks, structure-agnostic DNN-aided inference, and structure-oriented DNN-aided inference \cite{shlezinger2023model}.

In this paper, we focus on a DU network structure that is based on fractional programming (FP) due to its resilience to out-of-distribution (OOD) inputs in the formulated QoS-aware RSMA multiple-input-single-output (MISO) problem for wireless resource allocation. Note that the constraints in the QoS-aware projection problem are complex linear constraints instead of single equality constraint. To tackle this issue, the proposed DU approach incorporates learnable parameters into iterative algorithms, such as gradient descent (GD), allowing for convergence in fewer steps or layers. Moreover, violation control follows in each layer to rectify the layer level outputs and make outputs follow or approach to constraints. Computation complexity analysis shows that the proposed model is more efficient and more robust because of the simple and explicit structure. The major contributions are summarized as follows:
%Hence, DU method preserves the model’s resilience with OOD inputs and yields relatively reliable outputs.
%This proposed DU provides a new scheme to solve the QoS-aware projection problem in communication resource allocation to make violation control. 

\begin{itemize}

\item A QoS-aware RSMA multi-user MISO problem is formulated. FP is firstly applied to set a benchmark solution. The problem is then decoupled into unconstrained optimization and projection such that the projection gradient descent (PGD) algorithm can be applied. With learnable parameters added to the PGD, a DU scheme is employed to unfold the iterative PGD into several layers of deep learning neural networks.

\item A power factor projection framework is proposed based on \cite{NEURIPS2023_47547ee8} and \cite{cristian2023end} which reduces the projection problem search space dimension while maintaining the original layer level output beamformer directions. A special loss function is designed to reduce the QoS violation rate.

\item Experiments are conducted to demonstrate the performance and effectiveness of the proposed DU. We analyze the violation control, compare the DU approach with a data-driven deep learning model, and demonstrate the layer level out difference and similarity with traditional optimization algorithm. We also test the OOD on different channel SNR, maximum power, and QoS requirement. 
%comparison between traditional FP algorithm and proposed DU is given which shows the low-complexity, stability and reliability of proposed DU. Finally, the structural designing analysis for DU shows how the model-driven DU learn to optimize this QoS-aware problem. 
%These experiments given above confirm the proposed model performs well in performance, computation complexity and robustness and has potential in future communication networks.
\end{itemize}

The rest of the paper is organized as follows. Section \ref{sec:related_works} details the related works which covers RSMA resource allocation, data-driven deep learning in communication and model-driven deep learning in communication. Section \ref{sec:system_model} proposes the system model, formulated problem with model-driven method. In Section \ref{sec:DU_solution}, the deep unfolding is utilized on traditional algorithm with adding learnable parameters. Section \ref{sec:results} provides the numerical results for performance, OOD test, computation efficiency analysis and structural analysis. Finally, Section \ref{sec:conclusion} concludes this work.

\section{Related Works}\label{sec:related_works}
\subsection{RSMA Resource Allocation}

RSMA provides robustness and higher energy efficiency by broadcasting the common information, while freeing up bandwidth for private data in communication systems~\cite{yang2021optimization,clerckx2023primer,clerckx2016rate,clerckx2024multiple,xiao2023joint,dizdar2023rsma}. The authors in \cite{yang2021optimization} derived a closed-form solution for the optimal private beamformer and showed that RSMA has advantages in resource allocation tasks with a massive number of users, minimum rate demand of users, and low transmit power scenarios. Besides, it also provided a basic scheme for resource allocation in single-input single-output RSMA systems. The authors in~\cite{clerckx2023primer,clerckx2016rate,clerckx2024multiple} have demonstrated RSMA robustness, especially in scenarios with high latency sensitivity, which is essential in various communications environment. To reduce system latency, the lower computational workload is also required. Following this, there are some results focused on low complexity algorithms \cite{xiao2023joint,dizdar2023rsma}. Specifically, in \cite{xiao2023joint}, the authors proposed a low complexity algorithm by reducing the redundant constraints and pre-processing deployment region modeling. The authors in~\cite{dizdar2023rsma} designed zero forcing and maximum ratio transmission based methods to find closed-from expression solution for the RSMA max-min fairness problem and reduce computational complexity. To date, the majority of RSMA resource allocation problems are solved by traditional model-driven methods, which  consist of optimization theories such as convex or non-convex, or semi-convex ones.   

\begin{figure*}[ht]
    \centering
    \includegraphics[width=6in]{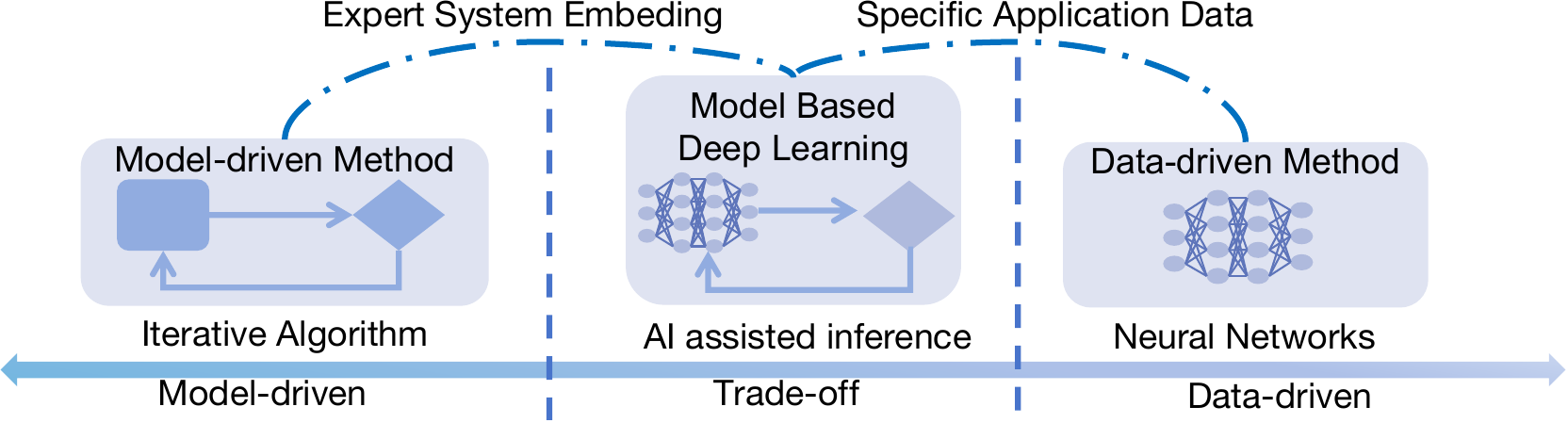}
	\centering
	\caption{The differences between model-driven method, data-driven method and model-based deep learning}
	\label{fig:DUprincipal}
\end{figure*}

\subsection{Data-driven Deep Learning in Communication}
Data-driven deep learning can train and get regression or diffusion to solve most of optimization problems with relatively lower complexity \cite{zhang2022deep,gao2022online,zhang2020deep}. Data-based deep learning or reinforcement learning studies are mostly applied on tackling communication problems \cite{huang2023deep,ye2019deep}. For example, reinforcement learning has been applied as Markov decision process to solve optimization problems~\cite{huang2023deep,ye2019deep}. Meanwhile, deep learning techniques such as GNNs, LSTMs, and CNNs are widely employed to address a multitude of optimization issues in communications. In particular, GNN has attracted considerable attention from scholars in communication-related research \cite{zhang2022learning,li2023graph,wang2023graph,xu2023distributed,zhang2023gnn}. The authors in~\cite{zhang2022learning} used GNN to train resource allocation networks, by considering the edge is weights and other status elements while edge is channel state information. GNNs are also effective in dynamic vehicular communication, federated learning, digital twin resource allocation optimizations, etc.~\cite{li2023graph,wang2023graph,xu2023distributed,zhang2023gnn}. Nonetheless, deep learning technology has been a essential part to optimize communication system performance and reduce computation consumption. 

\subsection{Model-based Deep Learning in Wireless Systems}
Model-based deep learning is developed based on interpreted model or a combination of  reasonable inference models \cite{shlezinger2023model}. Different DU approaches have been applied to symbol inference problem in communication by adding learnable parameters on GD polynomial to develop a specific optimizer~\cite{samuel2019learning,nguyen2020deep}. There are also more general studies on learning to optimize wireless resource allocation \cite{schynol2023coordinated,zhang2023joint,xia2023deep}. For example, the authors in~\cite{schynol2023coordinated} proposed a GCN-DU algorithm, which embeds GNNs into the weighted minimum mean-square error (WMMSE) algorithm to reduce computation complexity and uses model-based structure to restrict outputs within an acceptable region. The authors in~\cite{zhang2023joint} used Uzawa's method to reformulate original optimization problem and coupled the variable in constraints with objective function for DU network design. Other related works can be found applying DU on beamformer optimization resource allocation problem~\cite{zhang2022deep,xia2023deep}. The authors in~\cite{zhang2022deep} used WMMSE algorithm given in \cite{shi2011iteratively} to reformulate the original problem into a WMMSE-equivalent problem, then designed a DU network to solve the two-user RSMA beamforming problem. The authors in~\cite{xia2023deep} extended to a  more general scenario, where they proposed a FP-based DU framework to address weighted-sum-rate (WSR) problem in the RIS communication. However, these works can only provide approximation of an optimal solution, because of the limitations from variables appearing only in constraints and multi-complex nonlinear constraints. Moreover, most of the existing works focus only on the simple power constraints \cite{schynol2023coordinated,zhang2022deep}, although feasible results are required to meet both soft constraints and hard constraints.

%Therefore, most of resource allocation problems with more complex constraints make the results projection harder. 
Researches can be found focusing on a resource allocation problem with more complicated QoS-aware constraints~\cite{liang2019towards,eisen2020optimal,naderializadeh2020wireless,li2021multicell,alizadeh2023power}. For example, authors in \cite{liang2019towards} applied activation function (ReLu function) to find feasible results. Authors in~\cite{eisen2020optimal} provided a solution based on Lagrangian of original function, and trained to learn dual variables to punish the violation constraints. The authors in~\cite{naderializadeh2020wireless} used an unsupervised primal-dual counterfactual optimization approach with a slack variable to solve this problem. Authors in~\cite{alizadeh2023power} proposed deep explicit projection network with low complexity and no requirement of convexity. However, these data-driven deep learning models require large number of training data and complex learnable models. 

To summarize, as shown in Fig.~\ref{fig:DUprincipal}, model-based deep learning is a trade-off between model-driven algorithm (traditional algorithm) and data-driven methods (deep learning/ AI based methods). Model-based deep learning combines with the expert system and the inference ability learned from data.

% Notation: We use bold capital symbol as matrix, bold lowercase symbol as vectors, the lowercase symbol as scalar, $||.||$ as $l_2$ norm, $\text{Re}\{.\}$ as real part of function, $[.]^H$ as Hermite transpose, $[.]^T$ as transpose, $\mathcal{U}\{.\}$ as uniform distribution, $\mathcal{N}\{.\}$ as Gaussian distribution, $\Bar{z}$ as conjugate of $z$.

\begin{table*}[ht]
\centering
\caption{List of notations}
\label{notation_table}
\begin{tabular}{|c|c|c|c|} 
\hline
\textbf{Notations} & \textbf{Description}&\textbf{Notations} & \textbf{Description}\\
\cline{1-4} % 画一条横线从第二列到第三列
$\mathbf{h}_k$ & channel from BS to user $k$ & $\sigma_k^2$ & channel noise variance  \\
\hline
$\mathbf{x}_k$ & private signal transmitted to user $k$ & $\alpha_k$ & the sum rate weight for user $k$\\ 
\hline
${y}_k$ &  signal received by user $k$  & $\text{P}_c$ &  circuit energy consumption  \\ 
\hline
$\mathbf{x}_0$ & common signal transmitted to all users  & $\text{P}_\text{max}$ & total power consumption upper bound\\ 
\hline
$\mathbf{v}_k$ & private beamformer to user $k$ & ${r}_k$ & SINR lower bound for QoS requirement\\
\hline
$\mathbf{v}_0$ & common beamformer to user $k$ & $\mathbf{V}_k$ & exterior production of $\mathbf{v}_k$\\
\hline
$\Tilde{\mathbf{v}}_k$ & beamformers for user $k$ without projection& $z_k$ & fractional programming auxiliary variables\\
\hline
${R}_k^p$ & user $k$'s achievable private data rate & $l_{1,k},l_2,l_{3,k}$ & fixed learning rate\\
\hline
${R}_k^c$ & the common achievable rate for user $k$ & $\mathbf{o}_k$ & the derivative of penalty function over variable $\mathbf{v}_k$ or $\mathbf{v}_0$\\
\hline
${c}_k$ & the achievable common rate, decoded by user $k$ & $\bm{\theta}$ & learnable parameters \\
\hline
$n,N$ & $n$ is $n$-th layer, $\forall n \in \{1,2,...,N\}$ & $\mathbf{\beta}_k$ & the $k$-th channel interference DU expression\\
\hline
$i$ & the $i$-th iteration step & $\mathbf{\zeta}_k$ & the $k$-th learnable beamformer GD monomial\\
\hline
$\lambda$ & the constant penalty factor & $a_k^*,\Tilde{a}_k$ & the auxiliary variables to update $\mathbf{v}_k$ and $\mathbf{v}_0$\\
\hline
$\nabla \mathbf{v}_k,\nabla \mathbf{v}_0,\nabla R_k^c$ & the gradient for $\mathbf{v}_k,\mathbf{v}_0,\nabla R_k^c $& $\mathbf{\phi}$ & environment vector\\
\hline
bold lowercase & vector& bold uppercase & matrix\\
\hline
lower case & scalar& $\max(.)$& maximum value of expression \\
\hline
$\min(.)$& minimum value of expression & $\mathbf{A}^{-1}$ & pseudo-inverse or inverse of expression \\
\hline
$\bar{z}$& the conjugate of $z$ & $\mathbf{A}^{T}$, $\mathbf{A}^{H}$& transpose, Hermitian transpose\\
\hline
$\text{Re}\{z\}$& real part of $z$&$||.||_2$ & $l$-2 norm\\
\hline
\end{tabular}
\end{table*}

\section{System Model and Problem Formulation}\label{sec:system_model}
\subsection{System Model}

The studied downlink multi-user communication system consists of a single base station (BS) with $M$ antennas and $U$ single-antenna users. 
%In an RSMA system, data is split into common and private streams. The common stream carries the messages for all users (include traffic signals information, critical safety information) while the private stream provides individual data for each user. 
Denote $\mathbf{v}_0,\mathbf{v}_k \in \mathbb{C}^{M\times 1} $ as the beamforming vectors for common and private data streams, respectively. The BS transmit signal for common stream and user $k$ private stream are given by:
\begin{flalign}                     \mathbf{x}_0&=\mathbf{v}_0s[n], \\
        \mathbf{x}_k&=\mathbf{v}_ks_k[n], k\in\{1,2,..,U\},
\end{flalign}
where $s[n]$ and $s_k[n]$ are the common and dedicated data streams at time $n$, respectively. Both data streams are normalized, i.e., $E(s[n])^2 = E(s_k[n])^2 =1$. The received signal of user $k$ is:
\begin{flalign}
    {y}_k=\mathbf{h}_k^H\mathbf{x}_0 + \mathbf{h}_k^H\mathbf{x}_k+\sum_{j\neq k}^U \mathbf{h}_k^H\mathbf{x}_j + {n}_k,
\end{flalign}
where $\mathbf{h}_k \in \mathbb{C}^{M\times 1}$ is the channel between BS and user $k$, which follows circularly symmetric complex Gaussian distribution with power $h_k$, $\mathbf{h}_k \sim \mathcal{CN}(\mathbf{0},h_k^2\mathbf{I})$. ${n}_k\sim \mathcal{N} (0,\sigma_k^2) $ is AWGN noise. 
The $k$-th user's signal-to-interference plus noise ratio (SINR) is given below:
\begin{flalign}
    \text{SINR}_k = \frac{\mathbf{h}_k^H \mathbf{v}_k\mathbf{v}_k^H\mathbf{h}_k}{\sigma_k^2+\sum_{j \neq k}^U \mathbf{h}_k^H \mathbf{v}_j\mathbf{v}_j^H\mathbf{h}_k}. \label{SINR_def}
\end{flalign}
The achievable common stream data rate of user $k$ is:
\begin{flalign}
    &c_k=\log_2(1+\frac{\mathbf{h}_k^H \mathbf{v}_0\mathbf{v}_0^H\mathbf{h}_k}{\sigma_k^2+\sum_{j =1}^U \mathbf{h}_k^H \mathbf{v}_j\mathbf{v}_j^H\mathbf{h}_k}),\label{comm_stream}
\end{flalign}
while the private stream of user $k$ is:
\begin{flalign}
    &R_k^p=\log_2(1+\text{SINR}_k).\label{private_stream}
\end{flalign}
To guarantee common data stream at all users, the sum of user's common stream decoding rate must not exceed the lowest common data stream channel capacity
~\cite{clerckx2023primer,clerckx2016rate}, such that, 
\begin{flalign}
    \sum_{k=1}^U R_k^c\leq\min(c_k),
\end{flalign}
where $R_k^c$ is the common data rate allocated to $k$-th user. Without loss of generality, user 1 is assumed to have the lowest expected channel gain. Hence the sum of common stream rate upper bound~\cite{yang2021optimization} can be computed as:
\begin{flalign}
    \min(c_k) = \log_2(1+\frac{\mathbf{h}_1^H \mathbf{v}_0\mathbf{v}_0^H\mathbf{h}_1}{\sigma_k^2+\sum_{k=1}^U \mathbf{h}_1^H \mathbf{v}_k\mathbf{v}_k^H\mathbf{h}_1}).
\end{flalign}

\subsection{Problem Formulation}

The RSMA system is to achieve the maximum downlink WSR, as formulated in the optimization problem below:
\begin{subequations}
    \begin{flalign}
        \mathbf{P}_1 :&\max_{\mathbf{v}_k,R_k^c}   \text{WSR} =  \sum_{k=1}^U \alpha_k(R_k^c+R_k^p), \label{WSR_org}\tag{9}\\
        \text{s.t.}\
        &\sum_{k=0}^U \mathbf{v}_k^H \mathbf{v}_k+\text{P}_c \leq \text{P}_\text{max},\label{totalpowerconstraints}\\
        &\sum_{k=1}^U R_k^c\leq \min(c_k), \label{maxRcsum_org}\\
        &\text{SINR}_k \ge r_k\label{SINR_cons}\\
        &R_k^c\geq 0,\label{Rkc_exist}
    \end{flalign}\label{P1}%
\end{subequations}
where 
%Eq.~(\ref{WSR_org}) is the original problem in which the objective is to maximum the system WSR; 
$\alpha_k$ is the weight for user $k$; constraint Eq.~(\ref{totalpowerconstraints}) sets the total power consumption to be lower than BS power $\text{P}_\text{max}$, and $\text{P}_c$ is fixed circuit power consumption; constraint Eq.~(\ref{maxRcsum_org})guarantees common stream can be decoded by every user; constraint Eq.~(\ref{SINR_cons}) guarantees the users' private data transmission quality by setting SINR lower bound; and constraint Eq.~(\ref{Rkc_exist}) ensures the existence of common stream. Table \ref{notation_table} provides a summary of the notation used in this paper.

\subsection{Traditional Optimization Solution}
The iterative FP algorithm~\cite{shen2018fractional} is applied to set a benchmark solution to $\mathbf{P}_1$. In specific, the standard semi-definite relaxation (SDR) is applied, where let $\mathbf{V}_k=\mathbf{v}_k\mathbf{v}_k^H$. Besides, we define auxiliary variables: 
\begin{flalign}
  &z_k^*=(\sigma_k^2+\sum_{j\neq k}^U \mathbf{h}_k^H \mathbf{V}_j\mathbf{h}_k)^{-1}\mathbf{h}_k^H\mathbf{v}_k, \label{z_k_update}\\
  &z_0^{*}=(\sigma_k^2+\sum_{ k=1}^U \mathbf{h}_1^H \mathbf{{V}}_k\mathbf{h}_1)^{-1}\mathbf{h}_1^H\mathbf{{v}}_0^{i}\label{z0_update}.
\end{flalign}

Given initial feasible values for $\mathbf{v}_k$, $z_k^*$ and $z_0^*$ become constant. Let $\Phi_{0}=1+2\text{Re}\{\sqrt{\bar{z}_0\mathbf{h}_1^H \mathbf{{V}}_0\mathbf{h}_1 z_0}\}-\bar{z}_0(\sigma_k^2+\sum_{k=1}^U\mathbf{h}_1^H \mathbf{{V}}_k \mathbf{h}_1 )z_0$, and $\Phi_{k}=1+2\text{Re}\{\sqrt{\bar{z}_k\mathbf{h}_k^H \mathbf{{V}}_k\mathbf{h}_k z_k}\}-\bar{z}_k(\sigma_k^2+\sum_{j\neq k}^U\mathbf{h}_k^H \mathbf{{V}}_j\mathbf{h}_k )z_k$.
Then, from FP principle \cite{shen2018fractional}, $\mathbf{P}_1$ can be redefined as
\begin{subequations}
    \begin{flalign}
        \mathbf{P}_2 :&\max_{\mathbf{V}_k,R_k^c}  \sum_{k=1}^U \alpha_k \big(R_k^c+\log_2 (\Phi_k)\big), \tag{12}\\
        \text{s.t.}&\sum_{k=1}^U R_k^c\leq \log_2(\Phi_0),\label{maxRcsum}\\
        &\sum_{k=0}^U \text{Tr}(\mathbf{V}_k) +\text{P}_c \leq \text{P}_\text{max},\\
        &\frac{\mathbf{h}_k^H\mathbf{V}_k\mathbf{h}_k}{\sigma_k^2+\sum_{j \neq k}^U \mathbf{h}_k^H \mathbf{V}_j\mathbf{h}_k} \geq r_k,\\
        &R_k^c\geq 0.\notag
    \end{flalign}
    \label{P2}%
\end{subequations}
It can be readily shown that given $z_0^*$ and $z_k^*$, $\mathbf{P}_2$ is convex with respect to $\mathbf{v}_k$ and $R_k^c$. Therefore, it can be solved by well-known toolbox, such as \text{CVX} \cite{grant2014cvx}. Afterwards, $z_0^*$ and $z_k^*$ can be updated by the newly solved $\mathbf{v}_k$. \text{Algorithm \ref{algorithm_FP_1}} summarizes the iterative process.

%Perhaps add some description on why the traditional optimization is not good enough before intoducing DU method.

\begin{algorithm}
    \caption{FP Beamforming for WSR optimization}\label{algorithm_FP_1}
    \begin{algorithmic}
        \Require  $\mathbf{h}_k$, $\alpha_k$,  $P_0$, $P_c$, $P_\text{max}$, initial value of $z_0$, $z_k$, $\mathbf{v}_0$, $\mathbf{v}_k$ and $R_k^c$. Set counter $j=1 $ and convergence precision $\phi_p$. 
        \While{$|\text{WSR}_{j+1}-\text{WSR}_{j}| > \phi_p$}
            \State\textbf{Step 1} Update $\mathbf{V}_k$ and $\text{WSR}_{j}$ from $\mathbf{P}_2$ with ${z}_k^*$, ${z}_0^*$,
            \State\textbf{Step 2} Apply eigen decomposition on $\mathbf{V}_k$ to obtain $\mathbf{v}_k$,
            \State\textbf{Step 3} Update each $z_k^*$, $z_0^*$ by (\ref{z_k_update}) and (\ref{z0_update}), $j=j+1$.
        \EndWhile%
    \end{algorithmic}%
\end{algorithm}%

\begin{figure*}[!ht]
    \centering
    \includegraphics[width=6.2in]{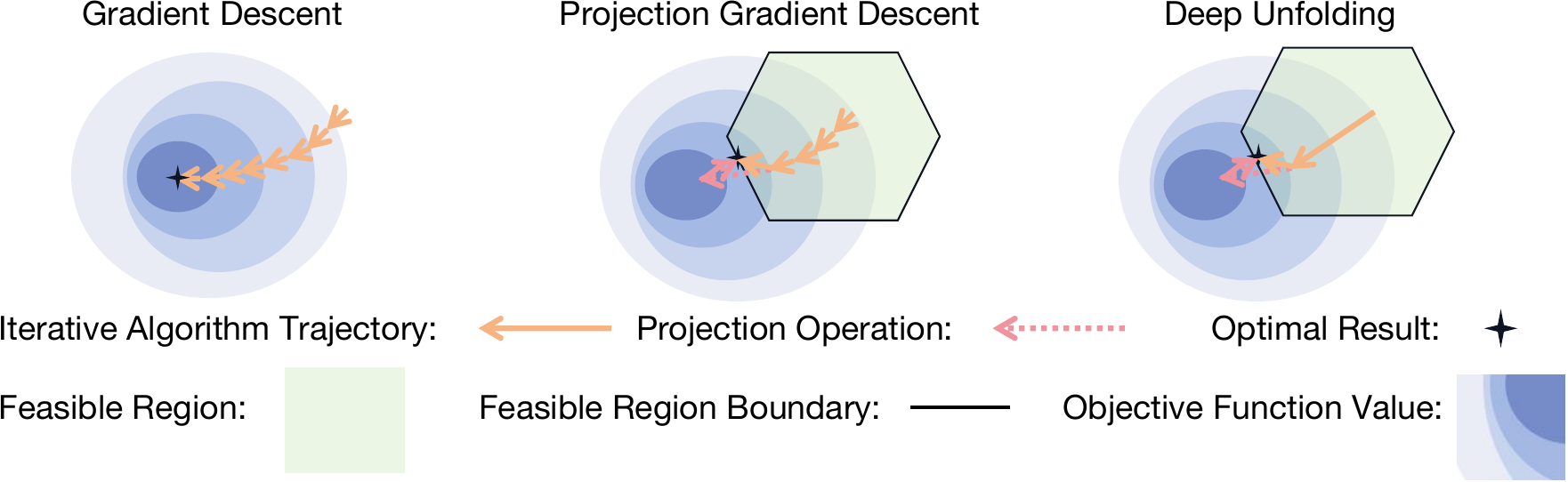}
 \centering
 \caption{The comparison of GD in unconstrained optimization, PGD and DU in constrained optimization}
 \label{fig:DUcomparison}
\end{figure*}
\section{Proposed DU Solution}\label{sec:DU_solution}

In this section, the DU design for this RSMA QoS-aware resource allocation problem is introduced. DU is based iterative optimization algorithm like GD while each iteration of iterative optimization algorithm is a layer in DU network. By adding learnable parameters in each layer, DU can perform well by training small scale of data with several layers. However, the learnable parameter may produce results that are out of feasible region. A projection problem is formulated to address this issue. %and \ref{fig:RSMA_DU_model}.

 As a model-based deep learning algorithm, DU's principal is given in Fig.~\ref{fig:DUcomparison}, in which it provides the differences among GD, PGD and DU. For GD and PGD algorithm, their trajectories are given by loss function gradient, while PGD can rectify the output value within feasible region by utilizing projection. Projection is a sub-problem to find the closest vector in feasible region to the output results. Therefore, PGD is able to solve constrained optimization problem. Similar to GD and PGD, DU has several layers and each layer can output the intermediate results as the input of next layer. As a consequence, the DU updates the results as same as GD and PGD till converge to optimal value. DU also has projection for layer as well, which guarantees the feasible output in whole procedure. The most significant difference between DU and PGD is DU's updating direction and learning rate is not only given by gradient but also the learnable parameters. The learnable parameters learn the trajectory experience from trainset data to find the better direction and learning rate to speed up convergence procedure. As a result, the DU can use only several layers to converge to a optimal value, much faster than PGD or other model-driven algorithm. In this section, we follow this principal and propose a specific framework to solve the complex vector projection in QoS constraint set. Based on the proposed DU, we apply end-to-end training to optimal DU performance on WSR and also the violation rate.

%\subsection{Iterative Updating Part}
\subsection{Variable Updating Part}
We start from the PGD based DU algorithm, which splits $\mathbf{P}_2$ into two parts: the GD on the objective function, a unconstrained problem and a projection to ensure all constraints are met. Since the derivative is directly based on $\mathbf{v}_k$ and $\mathbf{v}_0$ instead of SDR variable $\mathbf{V}_k$ and $\mathbf{V}_0$ in PGD algorithm, we use $\mathbf{v}_k$ and $\mathbf{v}_0$ in the following discussion.
Besides, dropping all constraints in $\mathbf{P}_2$ will lead to no gradient updates for $\mathbf{v}_0$. Therefore, we first reformulate $\mathbf{P}_2$ by adding a penalty function of constraint Eq.~(\ref{maxRcsum}) on the objective function, which couples gradient with $\mathbf{v}_0$. The unconstrained problem is formulated as:
\begin{flalign}
   \mathbf{P}_3:
    \max_{\mathbf{\Tilde{v}}_k,\Tilde{R}_k^c}   & \mathcal{F}=\sum_{k=1}^U \alpha_k\Big(\Tilde{R}_k^c+\log_2\big(\Tilde{\Phi}_k) \Big)\notag\\
    &-\lambda(\sum_{k=1}^U \Tilde{R}_k^c-\log_2(\Tilde{\Phi}_0)), 
    \label{P3}%
\end{flalign}%
where $\lambda > 0$ is the penalty factor; $\Tilde{\Phi}_{0}=1+2\text{Re}\{\sqrt{\bar{z}_0\mathbf{h}_1^H \mathbf{\Tilde{v}}_0\mathbf{\Tilde{v}}_0^H\mathbf{h}_1 z_0}\}-\bar{z}_0(\sigma_k^2+\sum_{k=1}^U\mathbf{h}_1^H \mathbf{\Tilde{v}}_k\mathbf{\Tilde{v}}_k^H \mathbf{h}_1 )z_0$; and $\Tilde{\Phi}_{k}=1+2\text{Re}\{\sqrt{\bar{z}_k\mathbf{h}_k^H \mathbf{\Tilde{v}}_k\mathbf{\Tilde{v}}_k^H\mathbf{h}_k z_k}\}-\bar{z}_k(\sigma_k^2+\sum_{j\neq k}^U\mathbf{h}_k^H \mathbf{\Tilde{v}}_j\mathbf{\Tilde{v}}_j^H\mathbf{h}_k )z_k$. When PGD updates $\mathbf{v}_k, \forall k$ in RSMA system, $\mathbf{v}_0$ can be updated in each iteration with the penalty function given in $\mathbf{P}_3$. Therefore, based on the unconstrained convex optimization in $\mathbf{P}_3$, the beamformers and common data rate $\mathbf{\Tilde{v}}_k^{i+1},\mathbf{\Tilde{v}}_0^{i+1}, \Tilde{R}_k^{c,{i+1}}$  can be  updated by $\mathbf{{v}}_k^i,\mathbf{{v}}_0^i, {R}_k^{c,i}$ following their GD direction. The iterative steps are given below: 
\begin{flalign}
    &\Tilde{R}_k^{c,i+1}=R_k^{c,i}+l_{1,k}\nabla{R}_k^{c,i}\notag,\\
    &\mathbf{\Tilde{v}}_0^{i+1}=\mathbf{v}_0^{i}+l_2\nabla\mathbf{{v}}_0^{i},\notag\\
    &\mathbf{\Tilde{v}}_k^{i+1}=\mathbf{v}_k^{i}+l_{3,k}\nabla\mathbf{{v}}_k^{i},\label{update_pgd}
\end{flalign}
where $i$ is the iteration index; $l_{1,k}$, $l_2$ and $l_{3,k}$ are fixed step size; and $\nabla{R}_k^{c,i},\nabla\mathbf{{v}}_0^{i}, \nabla\mathbf{{v}}_k^{i}$ are gradients (derivatives) of $\mathcal{F}$ with respect to $R_k^c$, $\mathbf{v}_0$, and $\mathbf{v}_k$, computed as
\begin{flalign}
    &\nabla{R}_k^{c,i} = \frac{\partial \mathcal{F}}{\partial \nabla{R}_k^{c,i}} = \alpha_k -\lambda, \forall k \neq 0,\label{GD_Rkc}\\
    &\nabla\mathbf{{v}}_0^{i}=\frac{\partial \mathcal{F}}{\partial \mathbf{v}_0^i}=\frac{2\lambda\bar{z}_0^{i}\mathbf{h}_1}{\Tilde{\Phi}_0^i\ln2},\label{V0_update_org}\\
    &\nabla\mathbf{{v}}_k^{i}=\frac{\partial \mathcal{F}}{\partial \mathbf{v}_k^i}=(\frac{\bm{\zeta}_k^i}{\Tilde{\Phi}_k^i}+\sum_{j\neq k}^U\frac{\bm{\beta}_{j,k}^i}{\Tilde{\Phi}_j^i}+ \mathbf{o}_{k}^i),\forall k \neq 0,\label{Vk_update_org}
\end{flalign}
where $\bm{\zeta}_k^i$, $\bm{\beta}_{j,k}^i$ and $\mathbf{o}_{k}^{i}$ are computed as
\begin{flalign}
    &\bm{\zeta}_k^i=\frac{\partial (2\alpha_k\text{Re}\{z_j^{i} \mathbf{h}_j^H \mathbf{{v}}_j^i\})}{\partial \mathbf{v}_k^i\ln2}={{2\alpha_k \bar{z}_k^{i} \mathbf{h}_k }}/\ln2,\\
    &\bm{\beta}_{j,k}^i=-\frac{\partial \alpha_k\bar{z}_j^{i}(\sigma_k^2+\sum_{l\neq j}^U\mathbf{h}_j^H \mathbf{{v}}_l^{i} \mathbf{{v}}_l^{i,H} \mathbf{h}_j )z_j^{i}}{\partial \mathbf{v}_k^i\ln2}\notag\\
    &~~~=-2 z_j^{i} \bar{z}_j^{i}\alpha_j\mathbf{h}_j  \mathbf{h}_j^H\mathbf{{v}}_k^{i}/\ln2,\forall k \neq 0,\\
    &\mathbf{o}_{k}^{i}=\frac{\partial\lambda\left(\sum_{k=1}^U -{R}_k^{i,c}+\log_2(\Tilde{\Phi}_0^i)\right)}{\partial \mathbf{{v}}_k^{i}}= \frac{-2 \lambda z_0^i \bar{z}_0^i\mathbf{h}_1\mathbf{h}_1^H \mathbf{{v}}_k^{i}}{\Tilde{\Phi}_0^i\ln2}.\label{Pn_update_org}
\end{flalign}

To apply DU on the GD method and guarantee that the variable updating is trainable, we add the learnable parameters in Eq. (\ref{V0_update_org}) and Eq. (\ref{Vk_update_org}) as follows:
\begin{flalign}
&\mathbf{\Tilde{v}}_0^{n+1}=\mathbf{v}_0^{n}+\Pi_1(\mathbf{{v}}_0^{n}),\label{V0_update_pi}\\
    &\mathbf{\Tilde{v}}_k^{n+1}=\mathbf{v}_k^{n}+\Pi_2(\mathbf{{v}}_k^{n}),\label{Vk_update_pi}
\end{flalign}
where $\Pi_1(\mathbf{{v}}_0^{n})$ and $\Pi_2(\mathbf{{v}}_k^{n})$ are the $n$-th layer in DU for  $\mathbf{v}_0$ and $\mathbf{v}_k$, $\forall n \in \{1,2,...,N\}$ which are computed as, 
\begin{flalign}
    &\Pi_1(\mathbf{{v}}_0^{n})=\ln2 (\bm{\phi}\mathbf{w}_0^n)\frac{\nabla\mathbf{{v}}_0^{n}}{\lambda},\\
    &\Pi_2(\mathbf{{v}}_k^{n})=(\bm{\phi}\mathbf{w}_k^n)\ln2
    \begin{bmatrix}\frac{\bm{\zeta}_k^n}{{\Tilde{\Phi}_k^n}},\frac{\bm{\beta}_{1,k}^n}{{{\Tilde{\Phi}_1^n}}},\frac{\bm{\beta}_{2,k}^n}{{{\Tilde{\Phi}_2^n}}},...\frac{\bm{\beta}_{j,k}^n}{{{\Tilde{\Phi}_j^n}}},..,\frac{\bm{\beta}_{U,k}^n}{{{\Tilde{\Phi}_U^n}}},\frac{\mathbf{o}_{k}^n}{\lambda}
    \end{bmatrix}\notag\\
    &\begin{bmatrix} \eta_k^n,\eta_j^{n,1},\eta_j^{n,2},..., \eta_j^{n,j},...,\eta_j^{n,U},\eta_k^{n,p}
    \end{bmatrix}^T, \forall k \neq 0,\forall j \neq k,\\
    &\bm{\phi}=[\alpha_1,\alpha_2,...,\alpha_U,\text{P}_\text{max}],
\end{flalign}
% & {\eta}_k^n = \bm{\phi}\bm{\eta}_k^n,{\eta}_j^{n,t} = \bm{\phi}\bm{\eta}_j^{n,t}, {\eta}_j^{n,p} = \bm{\phi}\bm{\eta}_j^{n,p},
where $\mathbf{w}_0^{n}, \mathbf{w}_k^{n}\in \mathbb{R}^{(U+1)\times 1},  {\eta}_k^{n,p}, {\eta}_j^{n,k}, {\eta}_k^{n} \in \mathbb{R}$; and $\bm{\phi}$ is the environment pattern vector, which is spliced by weights $\alpha_k$ and power consumption $\text{P}_\text{max}$. These learnable parameters are trained to dynamically adjust weights for beamformers at each iteration, thereby can converge quickly within a few steps. 

The design of learnable parameters is inspired by\cite{samuel2019learning,nguyen2020deep}, which adds parameters on each term in gradient polynomial to change the weights of updating in different directions. In this particular problem, the penalty function given above has penalty factor $\lambda$. Hence, the training process for parameter $\eta_k^p$ can be considered as the regression problem for learnable parameter $\eta_k^p$ to regress the Lagrangian multiplier value in each iteration step. The learnable parameter, $\mathbf{w}_k$ and $\mathbf{w}_0$, is applied to construct a linear transformation with $\bm{\phi}$ where this multiplication provides the adaptive updating direction which is related to these special parameters in environment victor. Therefore, the beamformers GD is able to be accelerated specifically with $\bm{\phi}$. 

\subsection{Constraint Projection Part}

To guarantee the updated variables in the feasible region of the original problem, we employ the power factor projection method to project $\mathbf{\Tilde{v}}_k$ to the closest value in its feasible set in $\mathbf{P}_2$ by reallocate beamformers' power. Since $R_k^c$ can be updated passively \cite{zhang2024model}, the projection algorithm for $\mathbf{\Tilde{v}}_k^{i}, \forall k \in \{0,1,...,U\}$ can be formulated firstly as
\begin{flalign}
    &\mathbf{P}_4:\label{P4} 
    \min_{\mathbf{{v}}_k} \mathcal{F}_1(\mathbf{{v}}_1),\mathcal{F}_2(\mathbf{{v}}_2),..., \mathcal{F}_k(\mathbf{{v}}_k)\\
    &\text{s.t. } \text{Eq.}~(\ref{totalpowerconstraints}), \text{Eq.}~(\ref{SINR_cons}),\notag%
\end{flalign}
where $\mathcal{F}_k(\mathbf{{v}}_k) = \frac{1}{2}{||\mathbf{{v}}_k-\mathbf{\Tilde{v}}_k||_2^2}$. Note that the solution for $\mathbf{P}_4$ requires a high-complexity algorithm. Therefore, according to the hard constraints projection framework given in \cite{NEURIPS2023_47547ee8}, we design a power projection algorithm which only focuses on the power allocation without changing beamformers' phase. As a result, this power projection problem reduces the search space significantly under the premise of satisfying the constraints, while ensuring a certain level of protection for the output of each layer of the deep learning model. Moreover, the iterative updating part for $\mathbf{v}_k$ represents the basis function and the power vector $\bm{\omega}$ is basis function weight in \cite{NEURIPS2023_47547ee8}. The power vector is defined as follows:
\begin{flalign}
    \bm{\omega}=&[{\omega}_1,{\omega}_2,...,{\omega}_k,{\omega}_0]^T,\notag\\
    \omega_k =& ||\mathbf{{v}}_k||_2^2.
\end{flalign}
Note that $\mathbf{{v}}_k$ are defined as variables. This power projection for beamformers reduces the computation complexity in Eq.~(\ref{P4}) while providing feasible results as long as the constraints are consistent. In sum, the approximate power projection problem is to find a feasible power vector that is closest to the original power vector given by GD. The problem is formulated as follows:
\begin{subequations}
\begin{flalign}
    \mathbf{P}_5:\label{P5}&
    \min_{\bm{\omega}} \frac{1}{2}||\bm{\omega}-\mathbf{a}||_2^2\tag{29}\\
    \textit{s.t.} \ &\sum_{k=0}^U \omega_k \leq \text{P}_\text{max}-\text{P}_c\label{power_unit_cons}\\ &{\omega_k\mathbf{h}_k^H \mathbf{\bar{v}}_k\mathbf{\bar{v}}_k^H\mathbf{h}_k}-r_k({\sigma_k^2+\sum_{j \neq k}^U \omega_k\mathbf{h}_k^H \mathbf{\bar{v}}_j\mathbf{\bar{v}}_j^H\mathbf{h}_k})\geq0,\label{sinr_unit_cons}
\end{flalign}
\end{subequations}
where $\bm{\omega}$ is a feasible power vector, and $\mathbf{a}$ is the original power vector given by GD, denoted as:
\begin{flalign}
    \mathbf{a}&=[a_1,a_2,...,a_k,a_0]^T,\label{a_vec}\\
    a_k &= ||\mathbf{\Tilde{v}}_k||_2^2,\\
    \mathbf{\bar{v}}_k&= {\mathbf{\Tilde{v}}_k}/{\sqrt{a_k}},\label{v_bar}\\
    \mathbf{v}_k&=\sqrt{\omega_k}\mathbf{\bar{v}}_k.\label{v_projected}
\end{flalign}
Besides, constraints Eq.~(\ref{power_unit_cons}) and Eq.~(\ref{sinr_unit_cons}) can be formulated in a constraint matrix $\mathbf{A}$, as described in Eq.~(\ref{A_matrix}).

\begin{figure*}[ht]
%\begin{strip}
%\centering
\begin{flalign}
\label{A_matrix}
\mathbf{A} = 
\begin{bmatrix} 
    \mathbf{h}_1^H\mathbf{\bar{v}}_1\mathbf{\bar{v}}_1^H\mathbf{h}_1& -r_1\mathbf{h}_1^H\mathbf{\bar{v}}_2\mathbf{\bar{v}}_2^H\mathbf{h}_1&...& -r_1\mathbf{h}_1^H\mathbf{\bar{v}}_k\mathbf{\bar{v}}_k\mathbf{h}_1&0\\
    -r_2\mathbf{h}_2^H\mathbf{\bar{v}}_1\mathbf{\bar{v}}_1^H\mathbf{h}_2& \mathbf{h}_2^H\mathbf{\bar{v}}_2\mathbf{\bar{v}}_2^H\mathbf{h}_2&...& -r_2\mathbf{h}_2^H\mathbf{\bar{v}}_k\mathbf{\bar{v}}_k\mathbf{h}_2&0\\
    \vdots&\vdots&\ddots& \vdots& \vdots\\
    -r_k\mathbf{h}_k^H\mathbf{\bar{v}}_1\mathbf{\bar{v}}_1^H\mathbf{h}_k& -r_k\mathbf{h}_k^H\mathbf{\bar{v}}_2\mathbf{\bar{v}}_2^H\mathbf{h}_k&...& \mathbf{h}_k^H\mathbf{\bar{v}}_k\mathbf{\bar{v}}_k\mathbf{h}_k&0\\
    -1&-1&\cdots&-1&-1
    \end{bmatrix}
\end{flalign}
%\end{strip}
\end{figure*}

% \noindent 
Then we reformulate the inequality as equality by adding real slack variable $\bm{\psi}=[\psi_1,\psi_2,...,\psi_U, \psi_0]^T$. The slack variable $\bm{\psi}$ is utilized to make inequality be consistent while the optimal results for projection is $\bm{\psi}=\mathbf{0}_{U+1}$, which means the output beamformers from each layer satisfy all the QoS constraints. Therefore, $\mathbf{P}_6$ is formulated as
\begin{subequations}
\label{P6}
\begin{flalign}
    \mathbf{P}_6:&
    \min_{\Tilde{\bm{\omega}}} f({\bm{\omega}})=\frac{1}{2}||\Tilde{\mathbf{\omega}}-\Tilde{\mathbf{a}}||_2^2\tag{35}\\
    \textit{s.t.} \ &\mathbf{A}\Tilde{\bm{\omega}}=\mathbf{b},\notag
\end{flalign}
\end{subequations}
where $\Tilde{\bm{\omega}} = [\bm{\omega};\bm{\psi}]$,  $\Tilde{\mathbf{a}} = [\mathbf{a};\mathbf{0}_{U+1}]$, and ${\mathbf{b}} = [r_1\sigma_1^2,r_2\sigma_2^2,\ldots,r_k\sigma_k^2,-\text{P}_\text{max}+\text{P}_c]^T$. Hence, constraint Eq.~(\ref{P6}) is formulated as equality constraints. 
%Matrix $\mathbf{A}$ is given as (\ref{A_matrix}). 

Please refer to Dykstra's projection method \cite{cristian2023end} for a solution to this quadratic programming problem. We can now project the solution given in DU part to the feasible region. For those results inconsistent to constraints, the Dykstra's projection method can project them to a trade-off point to all constraints. Therefore, the projection details in Eq.~(\ref{P6}) are formulated as follows:
\begin{flalign}
    \min_{\Tilde{\bm{\omega}}} \mathcal{L}(\Tilde{\bm{\omega}}) = \frac{1}{2}||\Tilde{\bm{\omega}}-\Tilde{\mathbf{a}}||_2^2-\lambda^T(\mathbf{A}\Tilde{\bm{\omega}}-\mathbf{b}),
\end{flalign}
which is the Lagrangian dual function of $\mathbf{P}_6$. To meet the KKT conditions, the derivative of $\mathcal{L}(\Tilde{\bm{\omega}})$ over $\Tilde{\bm{\omega}}$ is given as folows:
\begin{flalign}
    \Tilde{\bm{\omega}}-\Tilde{\mathbf{a}}+\mathbf{A}^T\lambda=0.\label{w_old}
\end{flalign}
Due to the complementary condition, we have $\mathbf{A}(\Tilde{\mathbf{a}}-\mathbf{A}^T\lambda)-\mathbf{b}=0$. Therefore, $\lambda$ can be represented as follows:
\begin{flalign}
    \lambda = (\mathbf{A}\mathbf{A}^T)^{-1}(\mathbf{A}\Tilde{\mathbf{a}}-\mathbf{b}). \label{lambda_rep}
\end{flalign}
By taking Eq.~(\ref{lambda_rep}) into Eq.~(\ref{w_old}), $\Tilde{\bm{\omega}}$ is formulated as
\begin{flalign}
    \Tilde{\bm{\omega}}=\Tilde{\mathbf{a}}-\mathbf{A}^T(\mathbf{A}\mathbf{A}^T)^{-1}(\mathbf{A}\Tilde{\mathbf{a}}-\mathbf{b}) \label{w_new}.
\end{flalign}
Since the power allocation for $\Tilde{\mathbf{v}}_k$ is given as Eq.~(\ref{v_projected}). Then
% \begin{flalign}
%     {\mathbf{v}}_k &= \sqrt{\Tilde{\omega}_k}{\bar{\mathbf{v}}_k},\notag\\
%     \Tilde{\bm{\omega}}&=[\Tilde{\omega}_1,\Tilde{\omega}_2,...,\Tilde{\omega}_k,\Tilde{\omega}_0,\mathbf{z}]^T, \label{v_projected}
% \end{flalign}
the corresponding projection for $\Tilde{R}_k^{c,n+1}$ for a common data rate allocation 
is computed as:
\begin{flalign}
    &R_k^{c,n+1}=\frac{\max(\Tilde{R}_k^{c,n},0)}{\sum_{k=1}^U\Tilde{R}_k^{c,n}}\min(c_k^{n+1}),\notag\\
    &=\frac{\max(\Tilde{R}_k^{c,n},0)}{\sum_{k=1}^U\Tilde{R}_k^{c,n}}\log_2\left(1+\frac{\mathbf{h}_k^H \mathbf{v}_k^{n+1}\mathbf{v}_k^{n+1,H}\mathbf{h}_k}{\sigma_k^2+\sum_{j\neq k}^U \mathbf{h}_k^H \mathbf{v}_j^{n+1}\mathbf{v}_j^{n+1,H}\mathbf{h}_k}\right).
\end{flalign}

When the number of iteration approaches to infinity, $R_k^{c*}$, $k\in\{k|\alpha_k= \max(\alpha_1,\alpha_2,...,\alpha_U)\}$, has asymptotic property to achieve upper bound, $\min (R_k)$, while other $R_k^c=0$. Therefore, projection step for $R_k^c$ is simplified as: 
\begin{equation}
R_k^{n+1,c} = 
\begin{cases} 
\min(c_k^{n+1}), & \text{if } \alpha_k = \max(\alpha_1,\alpha_2,...,\alpha_U),\\
0, & \text{otherwise}.
\end{cases}\label{R_kc_map}
\end{equation}

Based on the variables updating and projections above, the layer level designing is completed. The trainable parameters for DU networks' $n$-th layer are given as
\begin{flalign}
    \bm{\theta}^{n}=\{\mathbf{w}_0^{n,T},\mathbf{w}_k^{n,T}, {\eta}_k^{n}, {\eta}_j^{n,t} ,{\eta}_k^{n,p}\}.
\end{flalign}
The projection part is summarized in {Algorithm} \ref{projection_algorithm}.
\begin{algorithm}
    \caption{Projection Algorithm for QoS-Aware RSMA} \label{projection_algorithm}
    \begin{algorithmic}
        \Require the beamformers obtained by variable updating part:
        \State\textbf{Step 1} Update $\mathbf{a}$ by (\ref{a_vec}), $\mathbf{\bar{v}}_0$ and $\mathbf{\bar{v}}_k$ by (\ref{v_bar});
        
        \State\textbf{Step 2} Construct matrix $\mathbf{A}$ by (\ref{A_matrix}); 

        \State\textbf{Step 3} Update $\lambda$ by (\ref{lambda_rep});
        
        \State\textbf{Step 4} Update $\bm{\Tilde{\omega}}$ by (\ref{w_new});
        
        \State\textbf{Step 5} Update $\mathbf{{v}}_0$ and $\mathbf{{v}}_k$ by (\ref{v_projected});
         
        \State\textbf{Step 6} Update $ R_k^c$ by (\ref{R_kc_map}),
        \State\textbf{Step 7} Output results $\{\mathbf{v}_0,\mathbf{v}_k,R_k^c\}$.
    \end{algorithmic}
\end{algorithm}

%加一个图
\begin{figure}[ht] % 可选参数指定浮动位置
    \centering
    \includegraphics[width=3.2in]{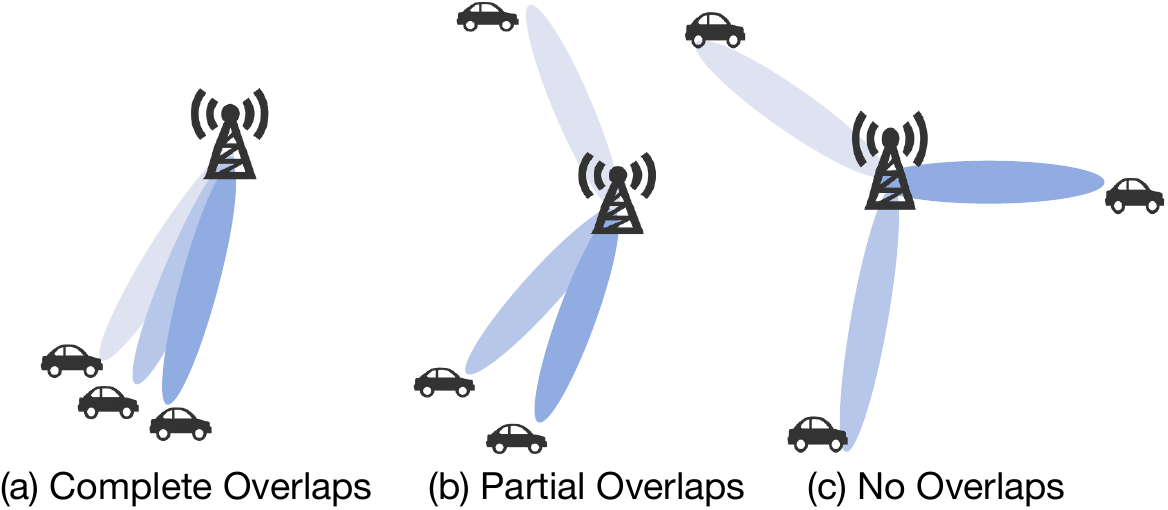} 
        \caption{Beams overlaps impact on power allocation projection}
        \label{fig:main_overlap}
\end{figure}

\subsection{Violation Control and Training Design}
When tackling the QoS projection, one of the most challenging tasks is reducing violation rate, i.e., when constraints are not met. Although we have a close-form expression algorithm given by projection  to meet the constraints, the beamformers' orthogonality may still fail to make power projection. When iterative updating part is processing, the beam overlap with each other can occasionally lead to higher interference within channels. However, {Algorithm} \ref{projection_algorithm} only considers power projection without phase projection. Hence, the phase for beam results in the inconsistent of constraints. Moreover, when it comes to complex resource allocation tasks, the violation rate is harder to be controlled. Fig.~\ref{fig:main_overlap} shows three different beam scenarios. In the scenario (a), the beam overlaps completely. In such occasion, whatever we change the beam power, the QoS can not be satisfied in feasible region. In the scenario (b), the beams have partial overlaps, in which some beamformers are able to be projected within feasible region. In scenario (c), the beams have no overlaps and all the beams are easily to meet QoS constraints. Moreover, the beamformers' initialization can also impact the violation control. As a result, we use conjugate of corresponding channel to do initialization for each beamformer where the learning task is easier for DU to control violation rate. 

Furthermore, to address violation control problem, we add penalty into loss function with a violation factor $\Xi^n$ for each layer, which is given as
\begin{flalign}
   \Xi^n =& \frac{1}{Q}\sum_{q=1}^Q\big(\text{ReLu}((\sum_{k=0}^U \text{Tr}(\mathbf{v}_k\mathbf{v}_k^H) +\text{P}_c)-\text{P}_\text{max})\notag\\
    & +\sum_{k=0}^U\text{ReLu}(r_k - \frac{\mathbf{h}_k^H\mathbf{v}_k\mathbf{v}_k^H\mathbf{h}_k}{\sigma_k^2+\sum_{j \neq k}^U \mathbf{h}_k^H \mathbf{v}_j\mathbf{v}_j^H\mathbf{h}_k})\big), \label{viiolation_design}
\end{flalign}
where $q$ is $q$-th sample in $Q$, which is the samples batch size. Since violation rate only counts the number of violation constraints while ignoring the violation level for each violation constraint. As a result, DU equally counts the violation punishment for all the infeasible beamformers, which makes training be harder than the proposed one. $\Xi^n$, the violation factor with $\text{ReLu}$ function, helps to smooth training procedure than using discrete violation rate. Besides, the violation factor can recognize the violation level, the critical violation makes higher punishment value in violation factor. Therefore, the loss function is given as
\begin{flalign}
    &\text{Loss}=\notag
    \\
    &\frac{-1}{QN}\sum_{q=1}^Q\sum_{n=1}^N\log_2(n+1)(\hat{\text{WSR}}_{q,n})+\sum_{n=1}^N\log_2(n+1)(\Xi^n),\label{loss_func}
\end{flalign}
where $\hat{\text{WSR}}_{q,n}$ is the WSR calculated by estimated beamformers for $n$-th layer and $q$-th samples
\begin{flalign}
    &{\hat{\text{WSR}}_{q,n}}=\notag\\
    &\sum_{k=1}^U \alpha_k^q\left({R}_{k,q}^{c,n}+\log_2(1+\frac{\mathbf{h}_{k,q}^H \mathbf{{v}}_{k,q}^n\mathbf{{v}}_{k,q}^{n,H}\mathbf{h}_{k,q}}{\sigma_{k,q}^2+\sum_{j\neq k}^U \mathbf{h}_{k,q}^H \mathbf{{v}}_{j,q}^n\mathbf{{v}}_{j,q}^{n,H}\mathbf{h}_{k,q}})\right).\label{estimated_wsr}%
\end{flalign}
In the loss function design, we add the penalty function with logarithmic weights similar to \cite{szegedy2015going}, the penalty in last layer output accounts the highest value. Besides, since this design is self-supervised, where the ground truth is not covered in loss function. Therefore, the generalization ability is better than the supervised one. Moreover, the violation punishment part still follows the logarithmic weights. Therefore, the DU networks can learn to output in feasible region with higher WSR. The proposed DU scheme is summarized in \text{Algorithm}.\ref{algorithm_DU}. 
\begin{figure*}[!ht]
    \centering
    \includegraphics[width=6.5in]{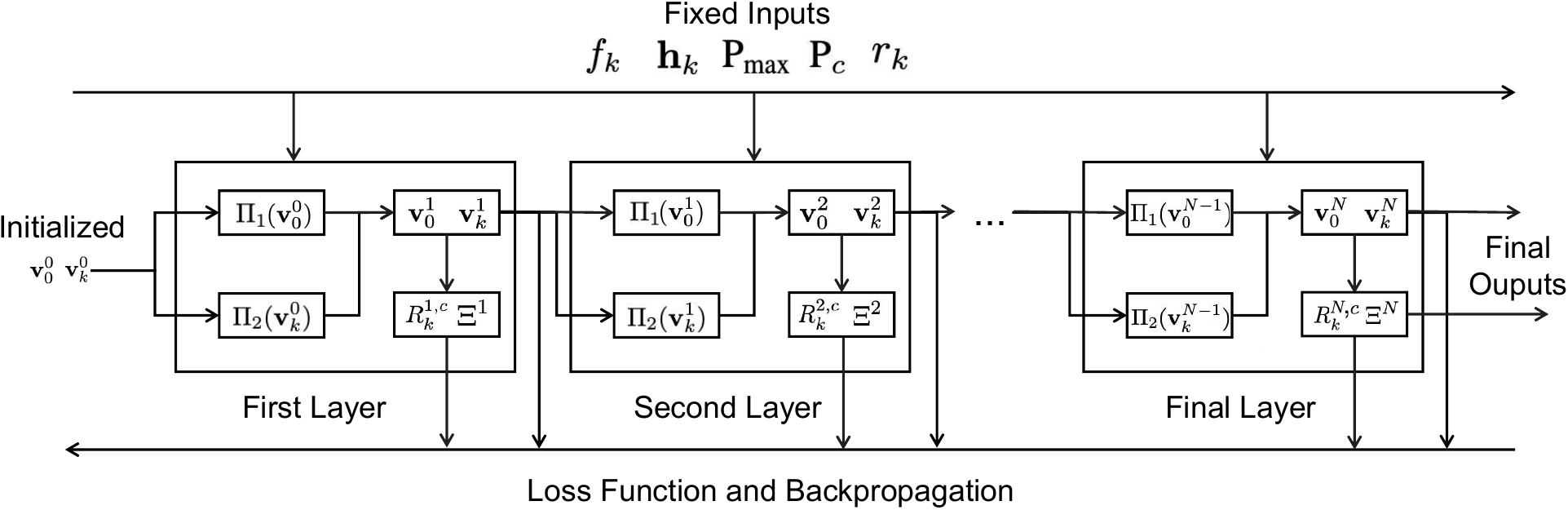}
	\centering
	\caption{Proposed DU networks structure overview}
	\label{fig:RSMA_DU_model}
\end{figure*}
\begin{algorithm}
    \caption{DU for FP-based WSR Optimization}\label{algorithm_DU}
    \begin{algorithmic}
        \Require  Initialize fixed parameters, learnable parameters, and set the layer number $N$, batch size, epoch, learning rate for optimizer in backward propagation,
        $\text{counter}=0$,
        
        \While{$\text{epoch} - \text{counter}>0$}\\
        $\text{counter}=\text{counter}+1$,\\
        \textbf{For each batch data:}\\
        
        \textbf{Forward Propagation Part:}
        
            \State Initial beamformers randomly, then for each layer:
            \State\textbf{Step 1} Update $z_k$, by (\ref{z_k_update}); Update $z_0$, by (\ref{z0_update}),
            \State\textbf{Step 2} Update $\mathbf{\Tilde{v}}_0$, by (\ref{V0_update_pi}); Update $ \mathbf{\Tilde{v}}_k$, by (\ref{Vk_update_pi}),
            \State\textbf{Step 3} Update $ \mathbf{{v}}_k$, $ \mathbf{{v}}_0$ and $ R_k^c$ by Algorithm \ref{projection_algorithm},
            \State\textbf{Step 5} Output results $\{\mathbf{v}_0,\mathbf{v}_k,R_k^c\}$,\\
        \textbf{Forward Propagation ends} \\
        \textbf{Backward Propagation Part:}
            \State\textbf{Step 6} Calculate the $\text{loss}$ given by (\ref{loss_func}),
            \State\textbf{Step 7} Backward Propagation, update $\bm{\theta}$ and optimizer.
            % \State\textbf{Step 8} Regulate parameters to be larger than $\epsilon$ by \\$\bm{\theta}^n=\max(\bm{\theta}^n,\epsilon$),\\
        \textbf{Backward Propagation ends}  
        \EndWhile
    \end{algorithmic}
\end{algorithm}
A graphic illustration of DU layers is shown in Fig.~\ref{fig:RSMA_DU_model}. %The forward propagation updating variables $\mathbf{v}_k$ and $\mathbf{v}_0$ by designed equation given in Eq.~(\ref{Vk_update_pi}) and Eq.~(\ref{V0_update_pi}). Then the projection of $\mathbf{\Tilde{v}}_k$, $\mathbf{\Tilde{v}}_0$ and $R_k^c$ is given by Algorithm \ref{projection_algorithm}. By repeating this procedure in each layer with different value of learnable parameters, DU can learn output approximate value of optimal beamforming strategy.

\section{Simulation Results}\label{sec:results}
% 基本参数介绍
The proposed DU is evaluated in three parts, including the DU convergence evaluation, comparison between model-based deep learning and data-driven deep learning, and the DU performance analysis. In all experiments, unless otherwise stated,  the settings are as follows: 3 users, $M=12$ antennas at the BS, 1600 epoches, learning rate is $0.01$, trainset size is $300$, batch size $bz = 200$, $N=4$, $\sigma^2 = 10^{-3}$, channel SNR is 15 $\text{dB}$, $\text{P}_\text{max}=$  $33\text{ dBm}$, $\text{P}_\text{c}=30\text{ dBm}$, and $r_k$ is set randomly from the distribution $|\mathcal{N}(0,1)|$.

\subsection{DU Convergence Evaluation}

Firstly, we evaluate the effectiveness of this DU networks by testing its performance in trainset and testset with different number of layers. The metrics are DU average sum ratio (ASR) and violation rate. ASR is the average ratio of optimal WSR for samples, computed as follows: 
%which is applied to evaluate the objective function performance, 
\begin{flalign}
    \text{ASR}=\frac{1}{Q}\sum_{q=1}^Q \frac{\hat{\text{WSR}}_{q,N}}{{\text{WSR}}_{q,N}^*},
\end{flalign}
where $\hat{\text{WSR}}_{q,N}$ is defined in (\ref{estimated_wsr}), ${\text{WSR}}_{q,N}^*$ is the optimal WSR given by the dataset generated by Algorithm \ref{algorithm_FP_1}.

\begin{figure}[ht!]
    \centering
    \includegraphics[width=3.3in]{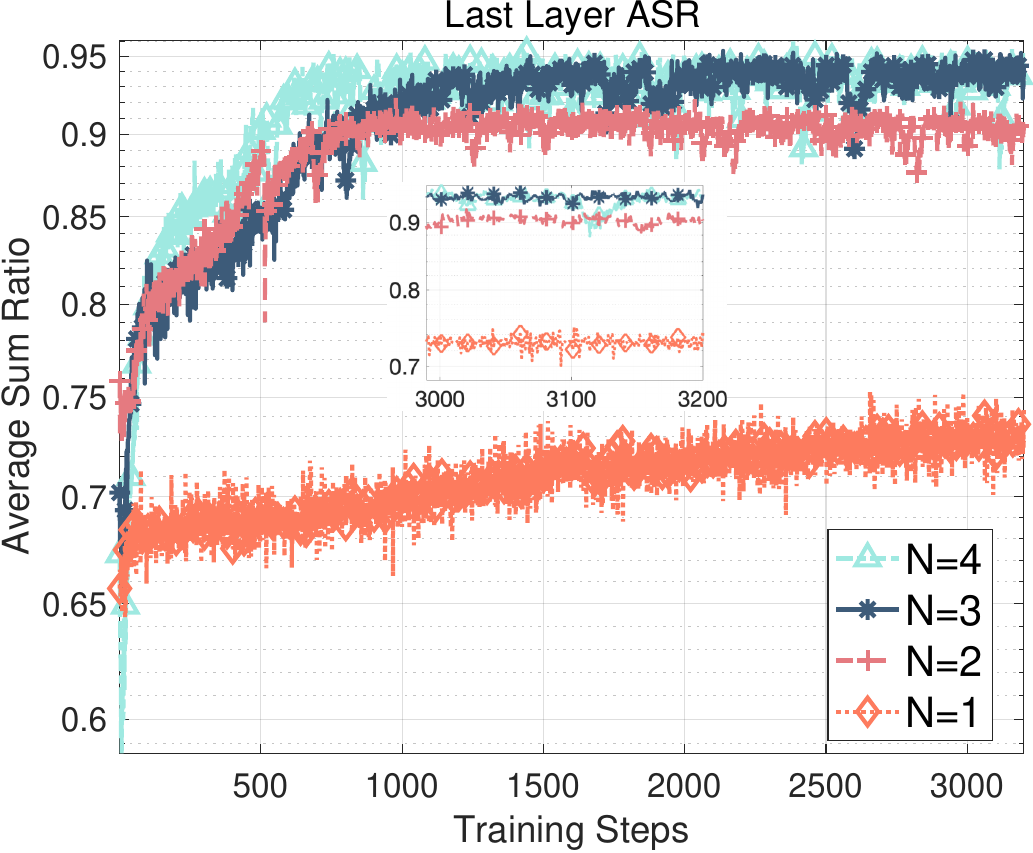}
	\centering
	\caption{Layer number impact on ASR performance}
	\label{ASR_layernum}
\end{figure}

\begin{figure}[ht!]
    \centering
    \includegraphics[width=3.2in]{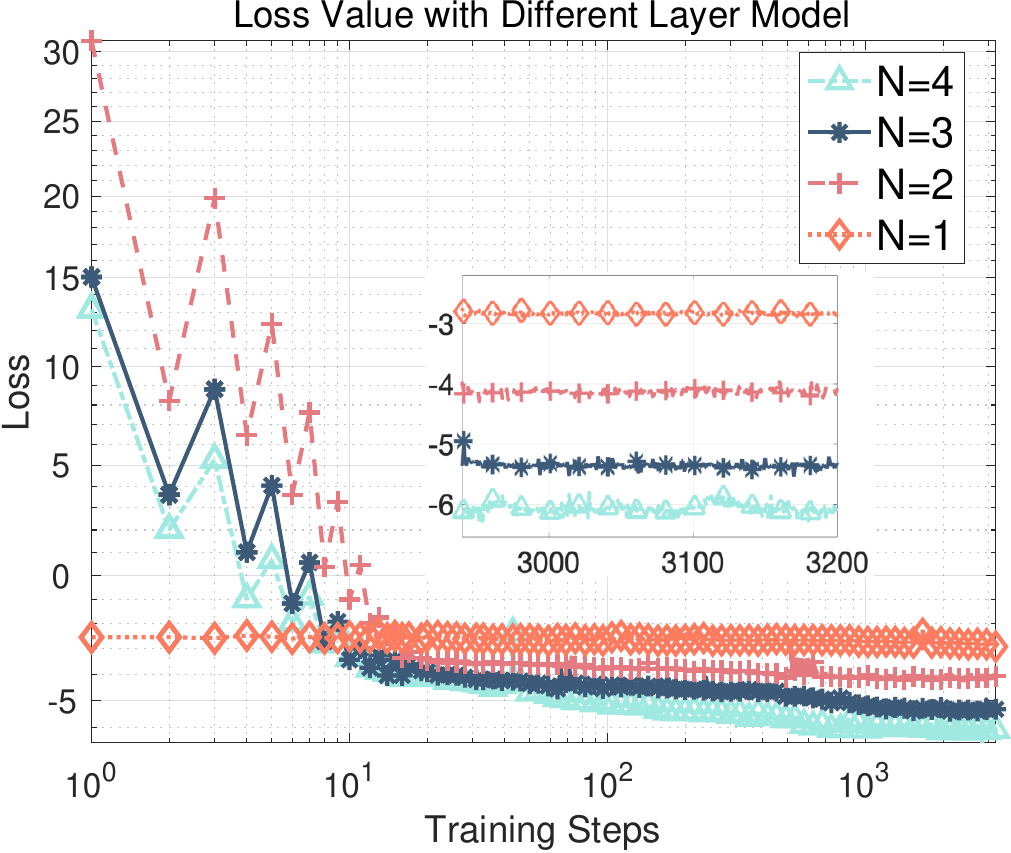}
	\centering
	\caption{Layer number impact on loss function performance}
	\label{Loss_layernum}
\end{figure}

Using the default parameter setting, 1-, 2-, 3-, and 4-layer DU are evaluated. Fig.~\ref{ASR_layernum} and Fig.~\ref{Loss_layernum} show the results of ASR and loss function, respectively. From the results we can see that the ASR increases as the number of layer increases. It is similar to an iterative optimization algorithm, e.g. GD, where more iterations usually yield a better result. When the layer number of DU increases to 3, the ASR and violation rate saturate at around $94\%$ and $0.024\%$ in testset, respectively. Detailed results are given in Table~\ref{ViolationRate_performancein_fristexp}. Moreover, the 3-layer DU has a similar ASR compared to the 4-layer DU. Nonetheless, ASR converges faster in the 4-layer DU than it does in the 3-layer DU, as shown in Fig.~\ref{ASR_layernum}. Meanwhile, although the 2-layer DU converges as fast as the 3-layer DU at the beginning, it can only achieve around $90\%$ ASR, whereas the 3-layer DU can achieve a higher ASR. It is worth noting that even the 1-layer DU model can reach $73.7\%$ ASR, which demonstrates that DU can direct the learning attention into a well-designed way to increase learning effectiveness and rapid convergence.

\begin{table}[ht!]
\centering
\caption{Testset performance for ASR}
\begin{tabular}{|l|c|c|c|c|c|} 
\hline
{Layer}&1&2&3&4 \\
\hline
{ASR (\%)}&$73.7$&$90.3$&$94.1$&$93.3$ \\
\hline
Violation Rate (\%)& $0.476$ & $0.095$ &  $0.024$& $0.024$ \\
\hline
\end{tabular}
\label{ViolationRate_performancein_fristexp}
\end{table}

Fig.~\ref{Loss_layernum} shows the original loss function value variation with different number of layers. Although the DUs with more layers have higher loss in the beginning, they can converge to a relatively lower level. This is because that more learnable parameters can lead to a more robust model. Also, more layers indicate more iterations in an optimization algorithm, which is consistent to this result. However, a DU with more layers needs more training steps to converge but with a better balance in both ASR and violation rate for its multi-objective loss function.

\subsection{Model-based vs. Data-driven Deep Learning}

\begin{figure}[ht!]
    \centering
    \includegraphics[width=3in]{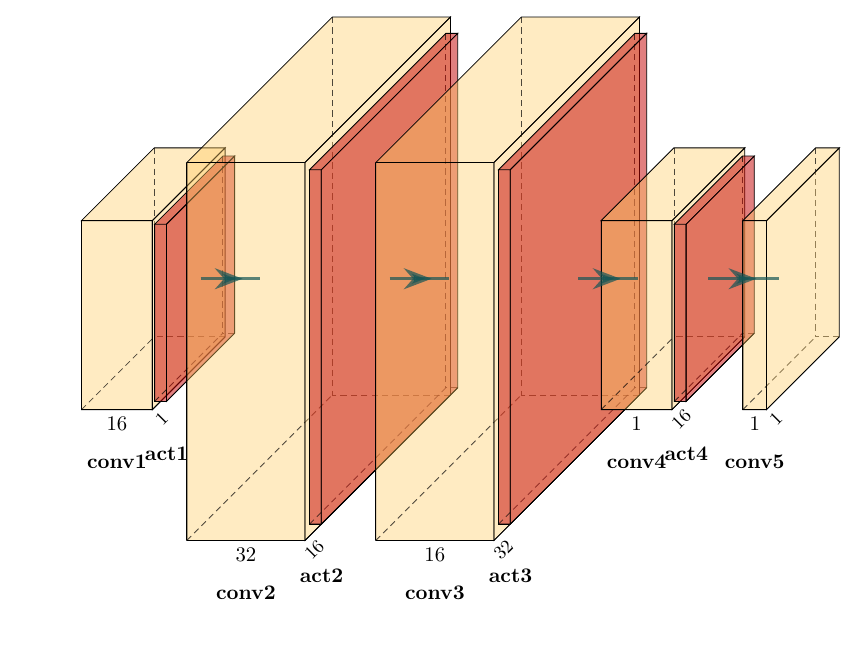}
	\centering
	\caption{Data-driven CNN structure}
	\label{CNNstruct}
\end{figure}

In this subsection, the effectiveness of model-based deep learning design is demonstrated. To show the ASR and violation control given by model-based DU network, we implement a data-driven deep learning approach for comparison. The deep learning approach is based on CNN, which has the structure given in Fig.~\ref{CNNstruct}. CNN has been commonly applied in different resource allocation problems~\cite{chen2019deep,lei2019learning,zhou2022deep,wang2020deep}. To ensure that the outputs have negative values, the sigmoid activation function is modified with a negative $1/2$ shift on the y-axis, followed by a multiplication with a positive constant $f$ to scale output range. In this experiment, we set $f=5$, and the layer parameters are given in Table \ref{CNN_parameters}.

\begin{table}[t]
\caption{CNN hyper-parameters}
\begin{tabular}{|l|c|c|c|c|c|} 
\hline
\textbf{Layer}&\textbf{conv1}&\textbf{conv2}&\textbf{conv3}&\textbf{conv4}&\textbf{conv5} \\
\hline
{Input ch. \#}&$1$&$16$&$32$&$16$&$1$ \\
\hline
Output ch. \#& $16$ & $32$ &  $16$& $1$ & $1$\\
\hline
Kernel \#& $3$ & $3$ &  $3$& $3$ & $3$\\
\hline
Padding& $1$ & $1$ &  $1$& $1$ & $0$\\
\hline
 Activation& $\text{Sigmoid}$ & $\text{Sigmoid}$ &  \text{Sigmoid}& \text{Sigmoid} & \text{None}\\
\hline
\end{tabular}
\label{CNN_parameters}
\end{table}

\noindent The input is $\mathbf{A}_{cnn}$, a matrix which is defined as 
\begin{flalign}
    \mathbf{A}_{cnn}&= [\mathbf{a}_{1};\mathbf{a}_{2};\mathbf{a}_{3};\mathbf{a}_{0}]\notag,\\
    \mathbf{a}_{1}&= [\text{Re}(\mathbf{h}_1),\text{Im}(\mathbf{h}_1),f_1,r_1],\notag\\
    \mathbf{a}_{2}&= [\text{Re}(\mathbf{h}_2),\text{Im}(\mathbf{h}_2),f_2,r_2],\notag\\
    \mathbf{a}_{3}&= [\text{Re}(\mathbf{h}_3),\text{Im}(\mathbf{h}_3),f_3,r_3],\notag\\
    \mathbf{a}_{0}&=({\mathbf{a}_{1}+\mathbf{a}_{2}+\mathbf{a}_{3}})/{3}.
\end{flalign}
As a consequence, the results given by this CNN model is $\mathbf{A}_{out}$, a tensor with dimension $[bz, 24, 4]$. 
Then the results are split as
\begin{flalign}
    \Tilde{\mathbf{v}}_1 = \mathbf{A}_{out}[bz,1:12,1]+j\mathbf{A}_{out}[bz,13:24,1]\notag\\
    \Tilde{\mathbf{v}}_2 = \mathbf{A}_{out}[bz,1:12,2]+j\mathbf{A}_{out}[bz,13:24,2]\notag\\
    \Tilde{\mathbf{v}}_3 = \mathbf{A}_{out}[bz,1:12,3]+j\mathbf{A}_{out}[bz,13:24,3]\notag\\
    \Tilde{\mathbf{v}}_0 = \mathbf{A}_{out}[bz,1:12,4]+j\mathbf{A}_{out}[bz,13:24,4]
\end{flalign}

\begin{figure}[ht] 
    \centering
    \begin{subfigure}[t]{0.4\textwidth}
        \centering
        \includegraphics[width=3.0in]{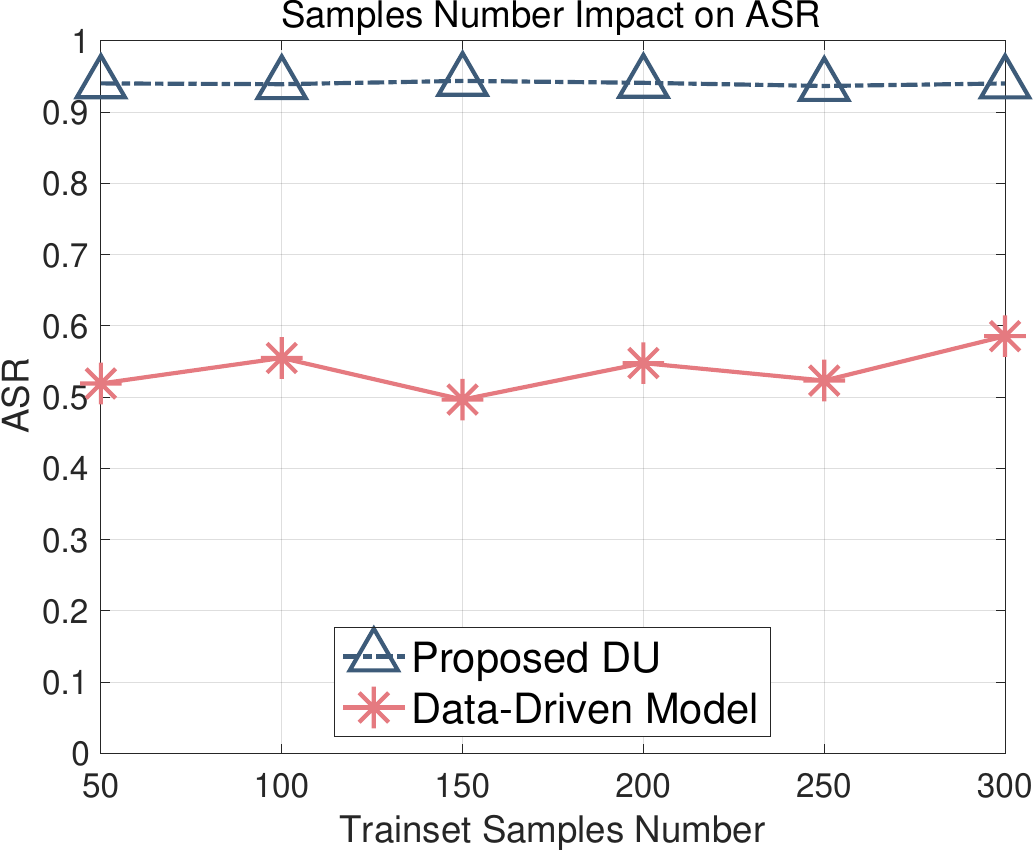} 
        \caption{ASR results}
        \label{ASR_trainsamples_num}
    \end{subfigure}\\
    %\hspace{10pt}
    \begin{subfigure}[b]{0.39\textwidth}
        \centering
        \includegraphics[width=3.0in]{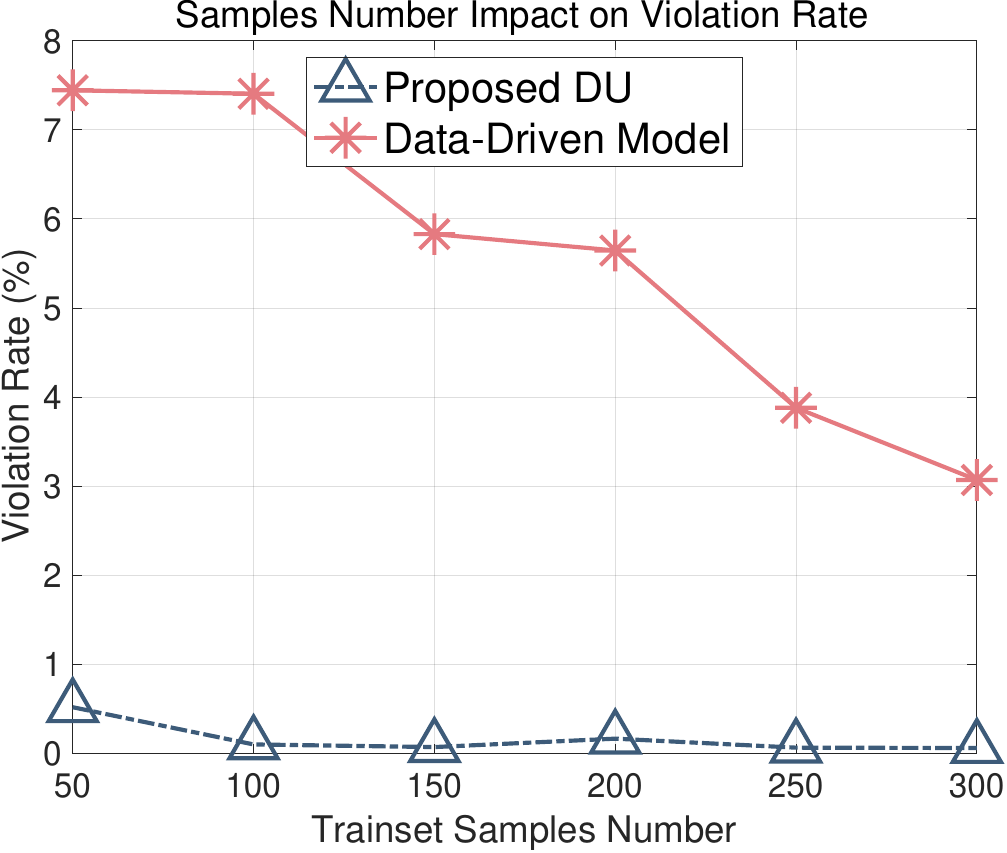} 
        \caption{Violation rates}
        \label{violation_trainsamples_num}
    \end{subfigure}    
    \caption{Trainset samples number vs. ASR and violation rate.}
    \label{train_num_diff}
\end{figure}

% \begin{figure}[ht!]
%     \centering
%     \includegraphics[width=3in]{hot_layer_ASR_layer1.pdf}
% 	\centering
% 	\caption{The ASR performance of each DU layer in testset}
% 	\label{hot_layer_ASR_layer}
% \end{figure}

\begin{figure*}[t] 
    \centering
    \begin{subfigure}[t]{0.28\textwidth}
        \centering
        \includegraphics[width=\textwidth]{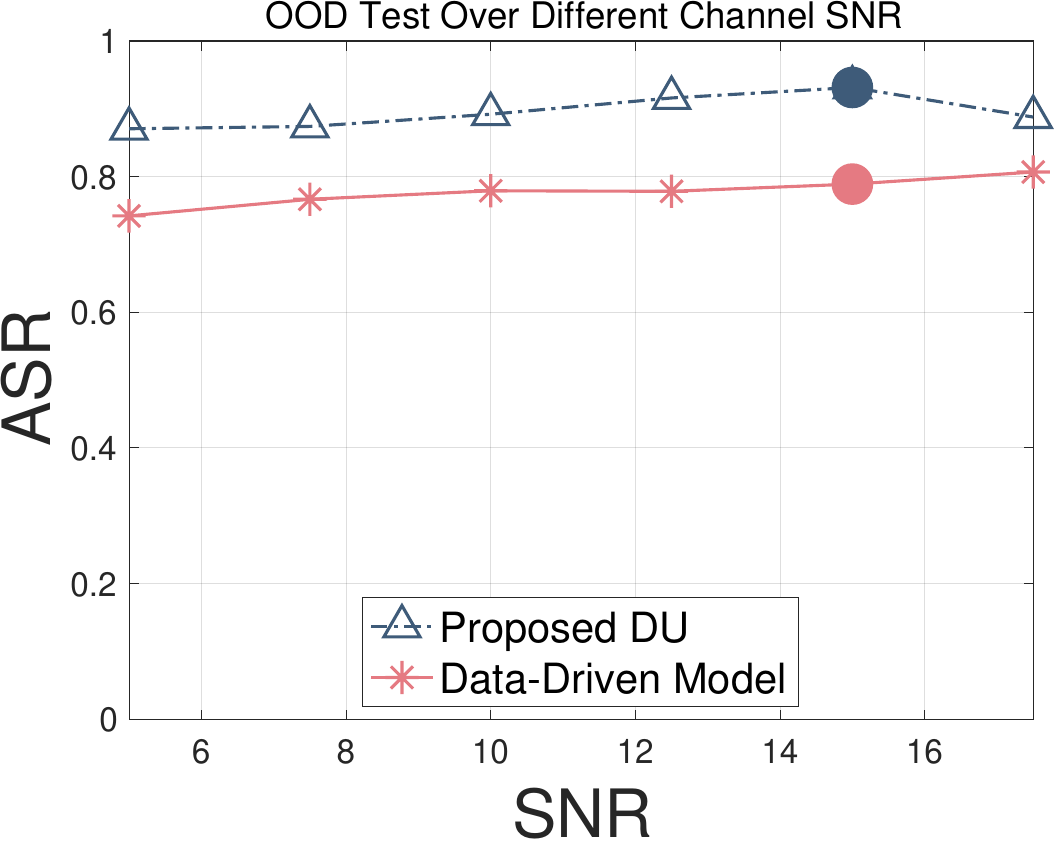} 
        \caption{OOD Test: SNR vs. ASR}
        \label{OOD_SNR_ASR}
    \end{subfigure}
    \hspace{10pt}
    \begin{subfigure}[t]{0.28\textwidth}
        \centering
        \includegraphics[width=\textwidth]{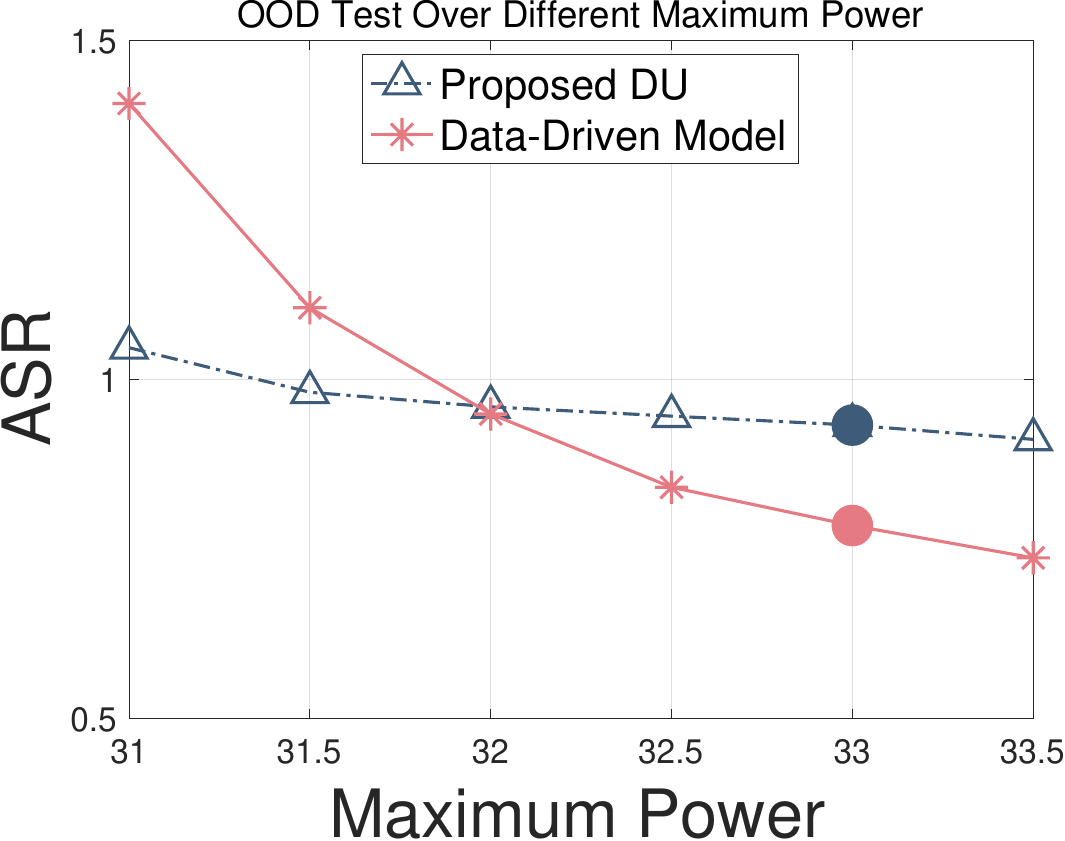} 
        \caption{OOD test: Power vs. ASR}
        \label{OOD_Power_ASR}
    \end{subfigure}
    \hspace{10pt}
    \begin{subfigure}[t]{0.28\textwidth}
        \centering
        \includegraphics[width=\textwidth]{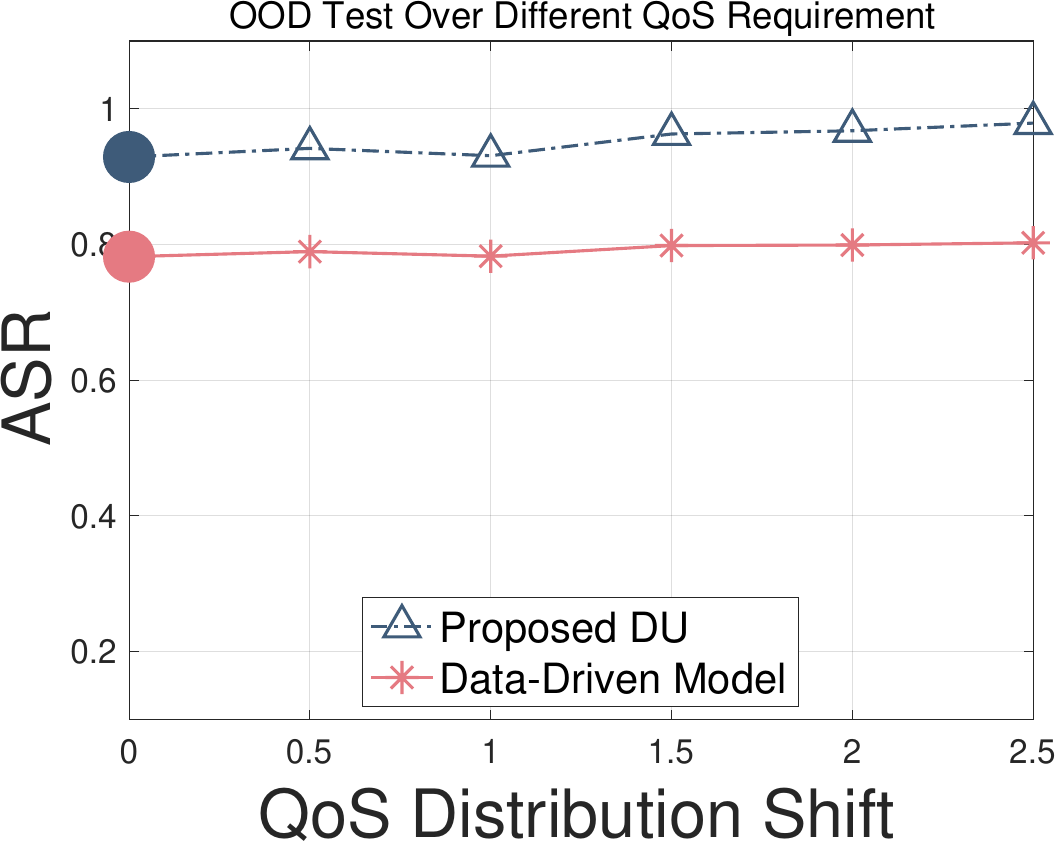} 
        \caption{OOD test: QoS vs. ASR}
        \label{OOD_QoS_ASR}
    \end{subfigure}
    \vspace{10pt}
    \begin{subfigure}[t]{0.28\textwidth}
        \centering
        \includegraphics[width=\textwidth]{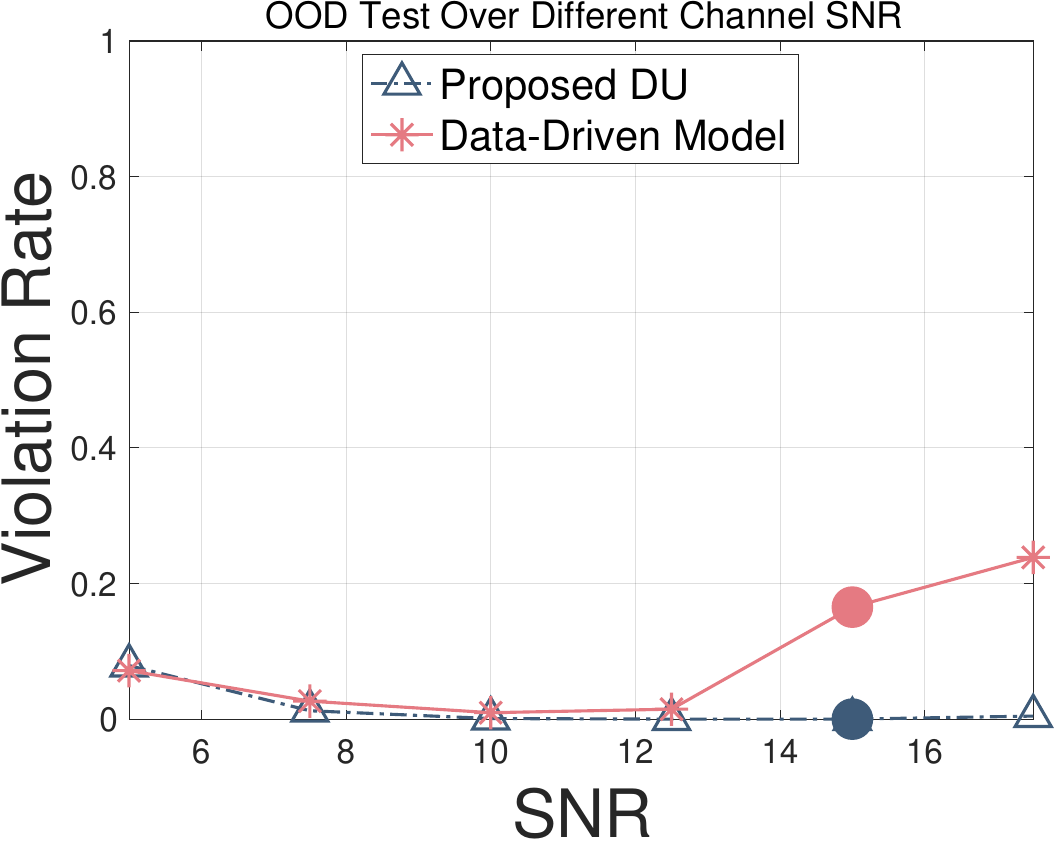} 
        \caption{OOD test: SNR vs. violation rate}
        \label{OOD_SNR_VR}
    \end{subfigure}
    \hspace{10pt}
    \begin{subfigure}[t]{0.28\textwidth}
        \centering
        \includegraphics[width=\textwidth]{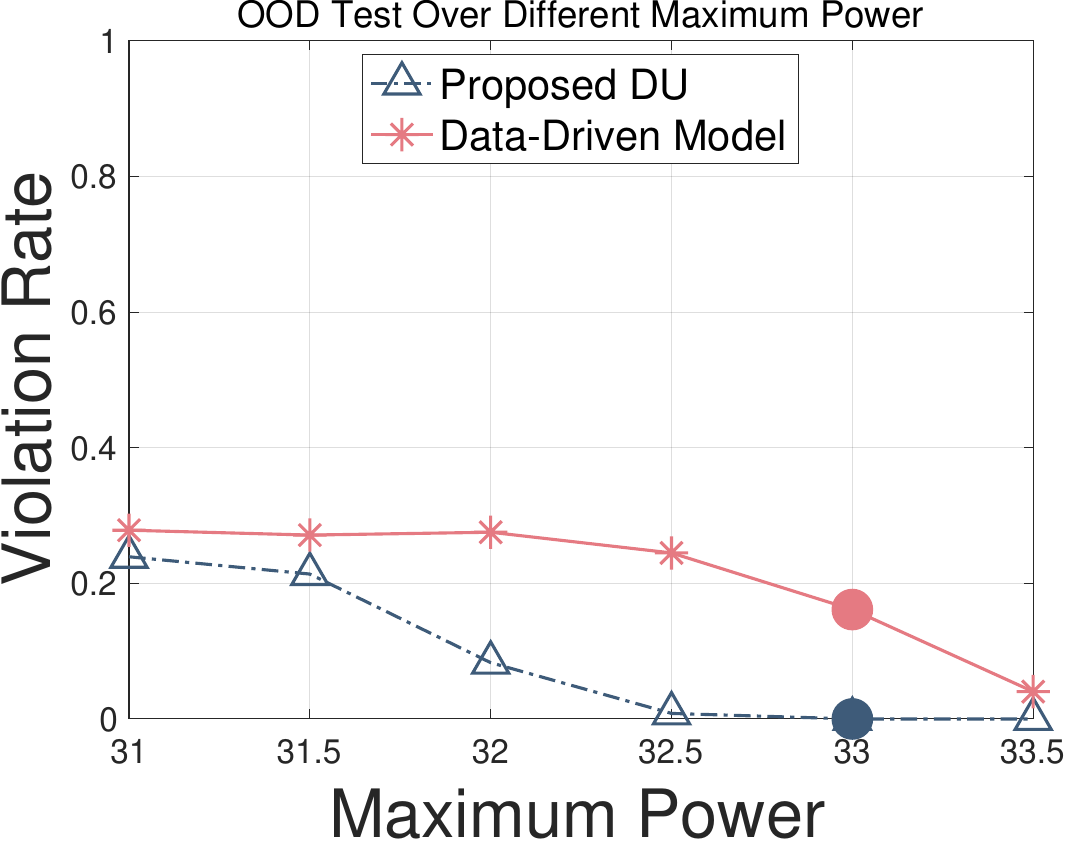} 
        \caption{OOD test: Power vs. violation rate}
        \label{OOD_Power_VR}
    \end{subfigure}
    \hspace{10pt}
    \begin{subfigure}[t]{0.28\textwidth}
        \centering
        \includegraphics[width=\textwidth]{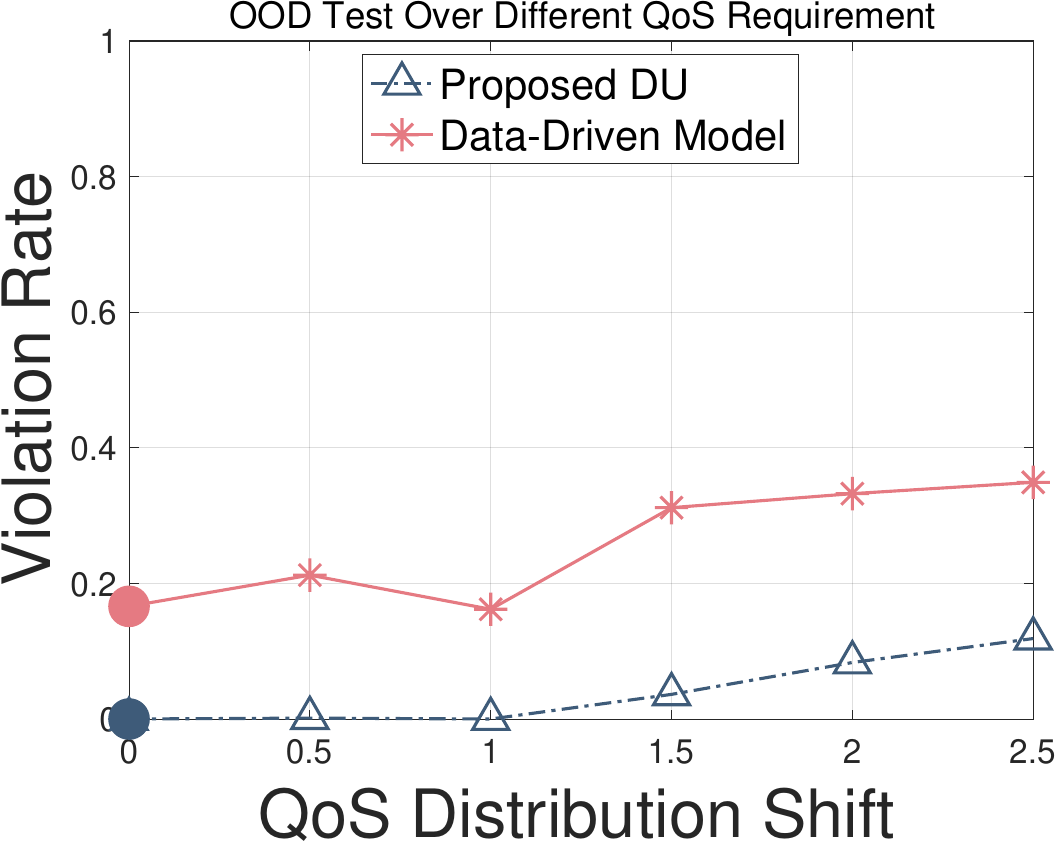} 
        \caption{OOD test: QoS vs. violation rate}
        \label{OOD_QoS_VR}
    \end{subfigure}
    \caption{OOD comprehensive performance evaluation}
    \label{OOD_test}
\end{figure*}

To ensure that the results given by CNN meet requirements, we apply the same projection method on this data-diven model. Therefore, the data-driven model outputs are also evaluated by ASR and violation rate. We use different number of trainset samples to train the data-driven, CNN model, and model-driven DU. Fig.~\ref{train_num_diff} shows the testset ASR and violation rate, respectively. For the proposed DU and CNN, we test 10 times for each number of trainset and obtain the average number for the ASR and violation rate. For the data-driven CNN model, the violation rate is $7.5\%$ at the beginning. And CNN's violation rate decreases with the increasing number of trainset samples to $3\%$ when it has 300 trainset samples. Besides, the CNN's average ASR is only around $60\%$. In comparison, the model-driven DU maintains its ASR around $92\%$ and violation rate less than $1\%$ even on 50 samples trainset. Moreover, the violation rate of proposed model-driven DU is close to $0$ from $100$ samples to $300$ samples which performs high confidence for its results.

Next, we evaluate the OOD performance, which tests the robustness of a model to new data distributions in the testset.  We set different OOD scenarios, including channel SNR OOD, maximum power OOD, and QoS requirement OOD. In each scenario, different distributions are applied to test the performance on well trained DU model. The OOD distribution for each scenario is given in Fig.~\ref{OOD_test}. For each OOD test, we mark the original parameter by solid circle. Fig.~\ref{OOD_SNR_ASR} and Fig.~\ref{OOD_SNR_VR}
show the results of channel SNR OOD test with $15$ dB as the standard channel SNR in the DU training dataset, while $5$ dB, $7.5$ dB, $10$ dB, $12.5$ dB, $17.5$ dB SNR scenarios are the OOD testset channel SNR scenarios. Note that the OOD data is not included in trainset which is tested directly by a well trained DU model. Therefore, this OOD test evaluates the generalization ability of the proposed DU. From $5$ dB to $17.5$ dB, the proposed DU has stable ASR and violation rate. Meanwhile, the violation rate of data-driven model in OOD dataset shows considerable variations. Especially on $5$ dB and $17.5$ dB dataset, data-driven model has a dramatic decreasing. Fig.~\ref{OOD_Power_ASR} and Fig.~\ref{OOD_Power_VR} show the results of maximum power OOD test. The maximum power ranges from $31$ dB to $33.5$ dB where the proposed DU maintains its ASR over $90\%$, while the data-driven model ASR decreases rapidly from $140\%$ to $72\%$. Note that under certain settings, ASR is over 100\% because of very high violation rate, which suggests an unreliable result. The corresponding violation rate also shows the proposed DU keeps better violation control than CNN. %which indicates DU has better power OOD tolerance. 
Fig.~\ref{OOD_QoS_ASR} and Fig.~\ref{OOD_QoS_VR} show the results of the OOD test with varying QoS requirement. The QoS requirement is by adding a constant number shift from $0$ to $2.5$ on the original $r_k$ distribution. Although both data-driven and proposed DU have stable ASR, the data-driven model has worse violation rate, which increases to $30\%$ with $1.5$ constant shift on QoS requirement.

\begin{figure}[ht!]
    \centering
    \includegraphics[width=3.2in]{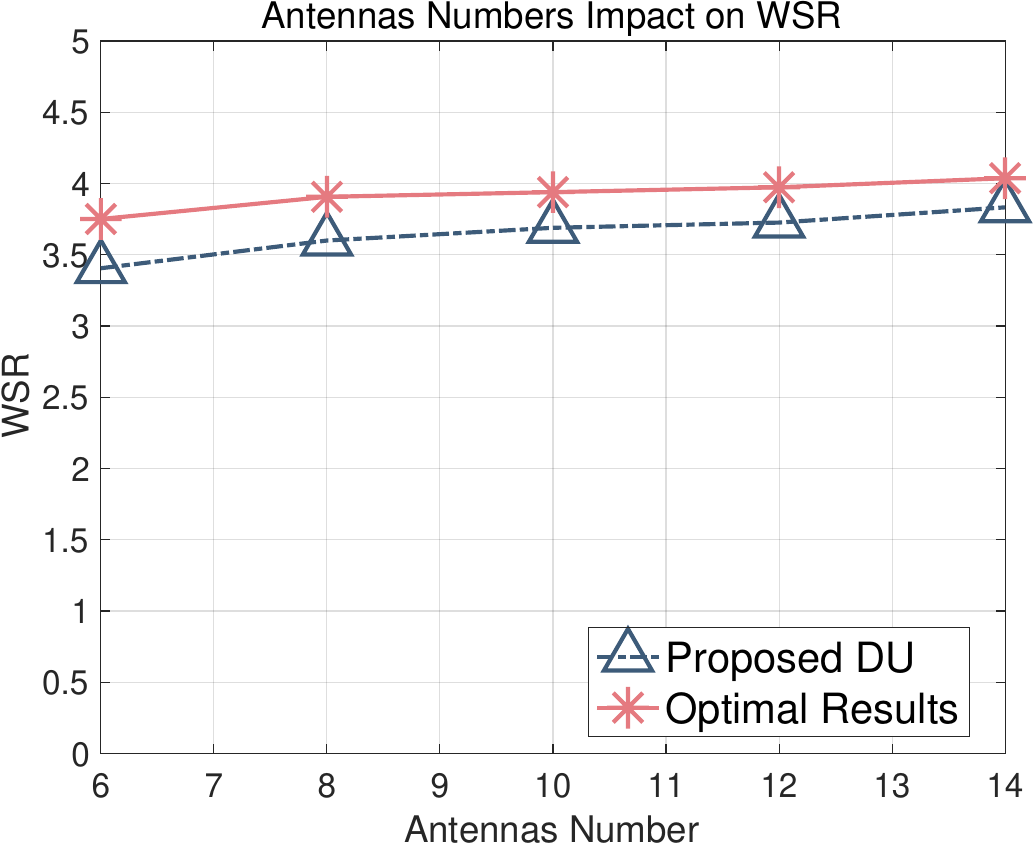}
	\centering
	\caption{Antennas number impact on WSR}
	\label{ATT_WSR}
\end{figure}

\subsection{DU Performance and Complexity Analysis}
In this subsection, the DU performance on different number of antennas, complexity and the loss function designing analysis are provided.
We first focus on evaluating the DU performance with different number of antennas. In particular, we test the antennas number of $6$, $8$, $10$, $12$, and $14$, respectively. The DU is trained on different dataset in this part. Fig.~\ref{ATT_WSR} shows the testset results. Generally speaking, the WSR performance maintains around $91\%$ while adding more antennas also helps achieve a higher ASR. Given the $6$-antenna settings, the ASR only achieves $90\%$. The ASR increases to $94\%$ when the transmitter has $14$ antennas.

\begin{figure}[ht!]
    \centering
    \includegraphics[width=3.2in]{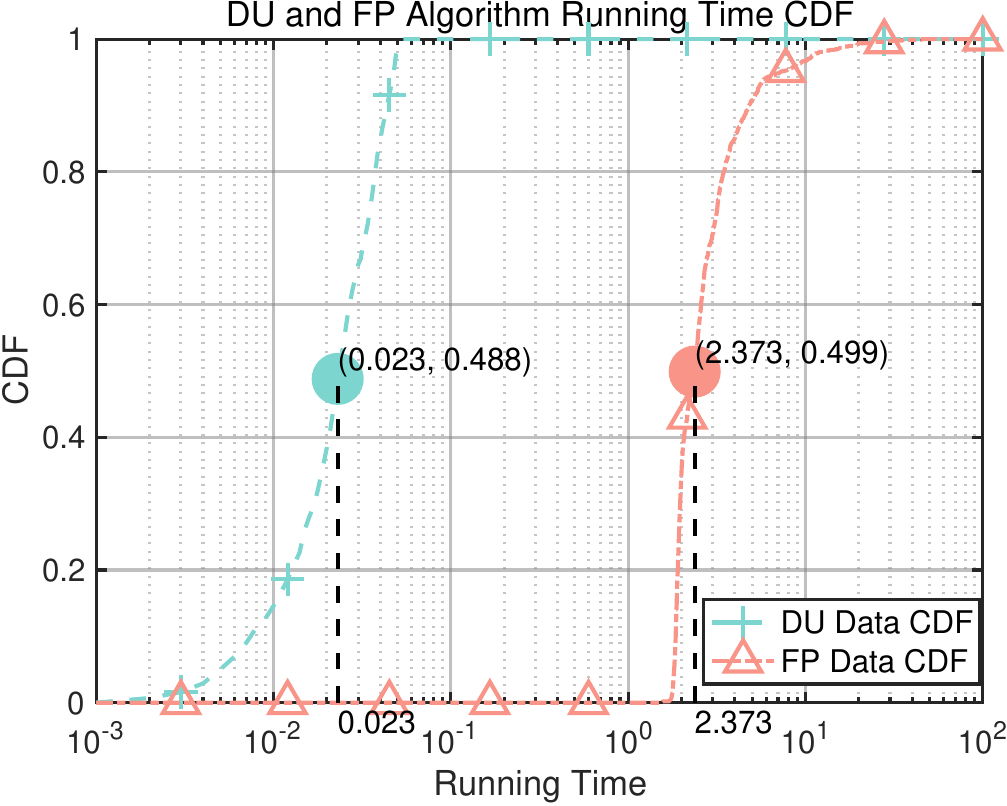}
	\centering
	\caption{Running time CDF of FP and proposed DU}
	\label{CDF_Computation_complexity}
\end{figure}

To show the superiority of the proposed DU algorithm in both computation and complexity, we compare its running time with the traditional FP algorithm. For a fair comparison, both DU and FP algorithms are implemented in Matlab. To avoid contingency of experiment, each algorithm is repeated 1000 times. Fig.~\ref{CDF_Computation_complexity} shows the corresponding cumulative distribution function (CDF) of DU and FP running time. We can see that the median running time for FP is 2.373 seconds, which is over 103 times more than the proposed DU at 0.023 seconds. Moreover, the proposed DU method calculates these results with more concentrated distribution while model-driven FP method's time ranges from 1 second to over 20 seconds. Statistically, the variance of DU data PDF and FP data PDF are around $3.9\times10^{-4}$ and $23.80$, respectively. Therefore, DU can support stringent time delay requirement tasks with reliable performance.

\begin{figure}[ht!]
    \centering
    \includegraphics[width=3.2in]{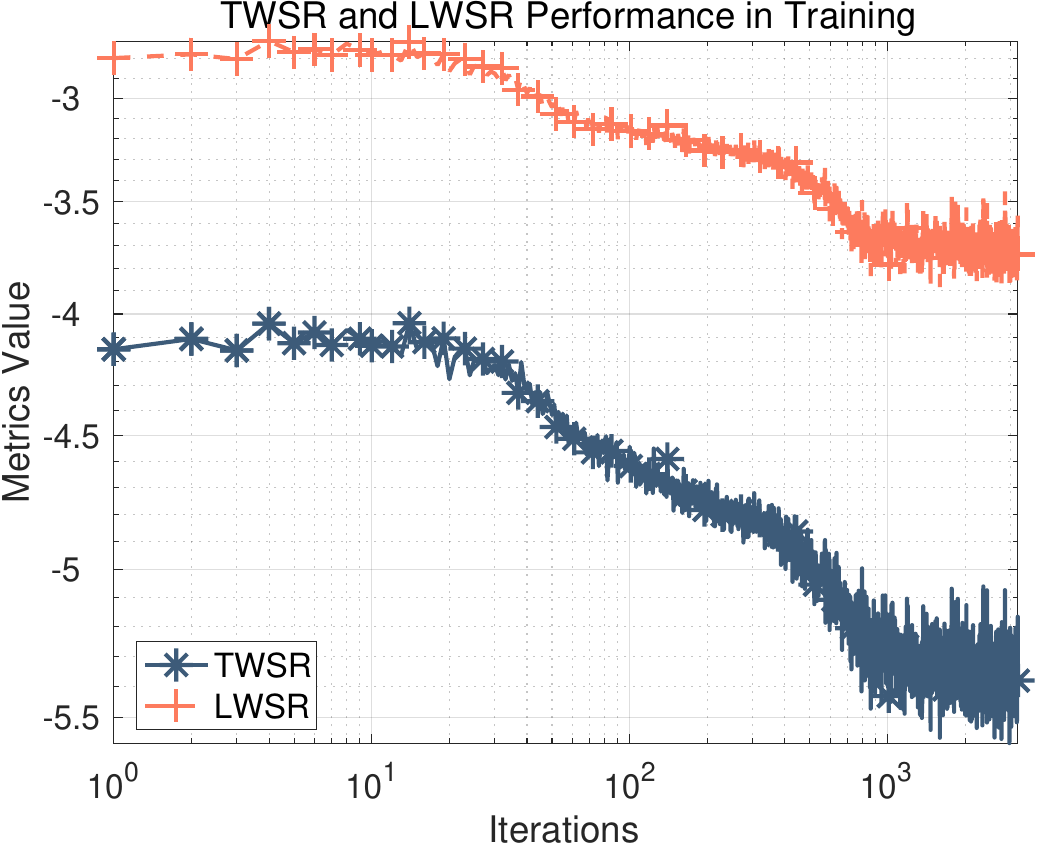}
	\centering
	\caption{Loss impact on TWSR and LWSR}
	\label{WSR_func_loss_relation}
\end{figure}

\begin{figure}[!ht]
    \centering
    \includegraphics[width=3.2in]{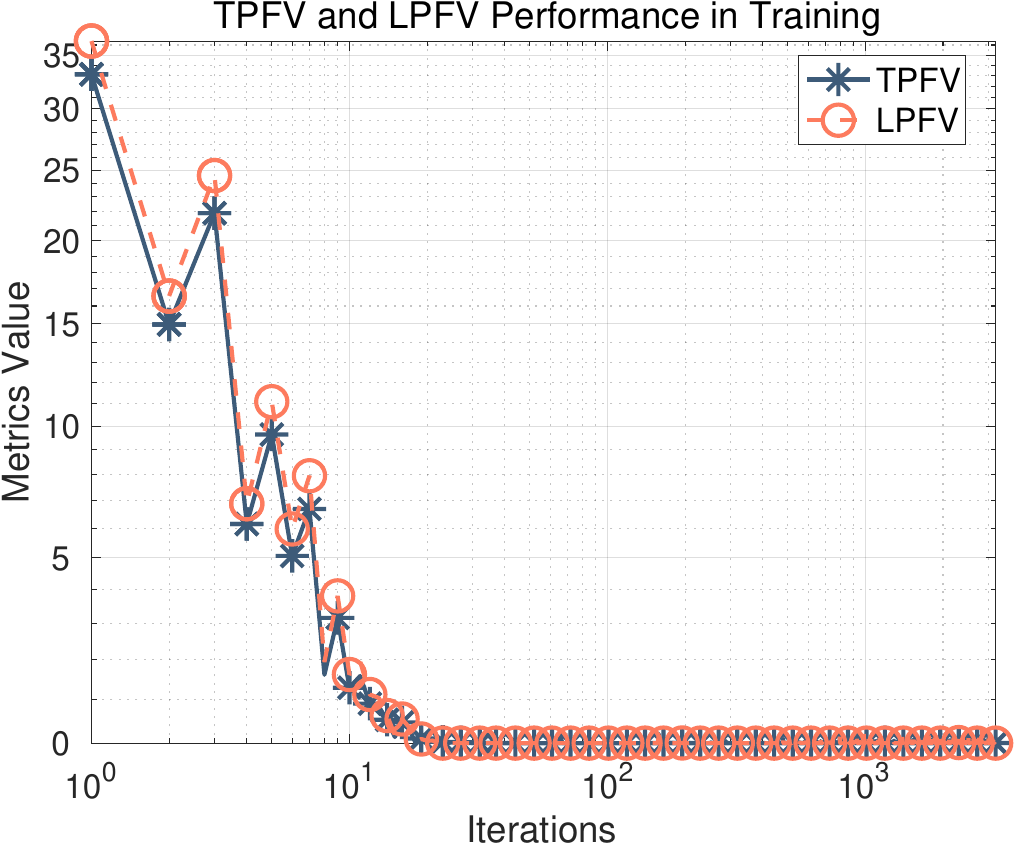}
	\centering
	\caption{Loss impact on TPFV and LPFV}
	\label{PFV_func_loss_relation}
\end{figure}

Lastly, we analyze the multi-modal loss design, which can be evaluated by total loss, total punishment function value (TPFV), last layer punishment function value (LPFV), total negative WSR (TWSR), and last layer negative WSR (LWSR). The definition of these metrics are given as follows:
\begin{flalign}
    &\text{TPFV}=\frac{1}{N}\sum_{n=1}^N\log_2(n+1)(\Xi^n),\notag
\end{flalign}
\begin{flalign}
    &\text{LPFV}=\log_2(N+1)(\Xi^N),\notag\\
    &\text{TWSR}=\frac{-1}{QN}\sum_{q=1}^Q\sum_{n=1}^N\log_2(n+1)(\hat{\text{WSR}}_{q,n}),\notag\\
    &\text{LWSR}=\frac{-1}{Q}\sum_{q=1}^Q\log_2(N+1)(\hat{\text{WSR}}_{q,N}),
\end{flalign}
where TPFV and TWSR are the metrics of average punishment function value and average negative WSR value in loss function for all layers. While the same time LPFV and LWSR are punishment function value and average negative WSR value given by the last layer output, respectively. Fig.~\ref{WSR_func_loss_relation} shows the results of loss function value, TWSR and LWSR. Because of over violation, both TWSR and LWSR increase to high levels in the beginning of a training process. The initialization has no awareness of violation rate and the punishment function value dominants the loss function. Therefore, the DU learns to reduce violation rate by giving up part of WSR performance. But the violation rate decreases immediately within a few steps. On the contrary, the WSR part in loss function dominates the loss value. Otherwise, the designed loss function shares the same descent direction with LWSR which is highly related to ASR and LWSR. 
Fig.~\ref{PFV_func_loss_relation} shows the connection between TPFV and LPFV. Both curves decrease within 25 iterations. This is caused by the extremely high punishment in the beginning, when the DU learns to restrict the impact from violation rate firstly. LPFV is similar to TPFS, which indicates that the violation control performs well not only in the last layer, but also in all the other layers.
%Therefore, the Fig.~\ref{WSR_func_loss_relation} and Fig.~\ref{PFV_func_loss_relation} prove the effectiveness of proposed loss function on ASR and violation control. 

%The results are given in Fig.~\ref{WSR_func_loss_relation}, in which we find the TWSR relative smoothly descent in training time while the LWSR generally follows the same where the DU reassigns the output weights on other layers to optimize DU performance.

\section{Conclusions}\label{sec:conclusion}

In this paper, we proposed a DU approach and a unique power factor projection method to solve the resource allocation problem in the QoS-aware RSMA multi-user communication system. Experiments were conducted to show the WSR and violation control stability under different settings, include the OOD scenarios. Comparison to a data-driven CNN model demonstrated that the proposed DU can be trained with smaller dataset while performing similarly to the model trained by a larger dataset. The computation complexity and structural analysis for DU further demonstrated that the proposed DU  model-based deep learning is more effective, less complex, and more stable than the traditional FP algorithm, as well as data-driven deep learning approaches for future communication systems. 

\bibliographystyle{IEEEtran} 
\bibliography{lib}

\begin{thebibliography}{10}
\providecommand{\url}[1]{#1}
\csname url@samestyle\endcsname
\providecommand{\newblock}{\relax}
\providecommand{\bibinfo}[2]{#2}
\providecommand{\BIBentrySTDinterwordspacing}{\spaceskip=0pt\relax}
\providecommand{\BIBentryALTinterwordstretchfactor}{4}
\providecommand{\BIBentryALTinterwordspacing}{\spaceskip=\fontdimen2\font plus
\BIBentryALTinterwordstretchfactor\fontdimen3\font minus \fontdimen4\font\relax}
\providecommand{\BIBforeignlanguage}[2]{{%
\expandafter\ifx\csname l@#1\endcsname\relax
\typeout{** WARNING: IEEEtran.bst: No hyphenation pattern has been}%
\typeout{** loaded for the language `#1'. Using the pattern for}%
\typeout{** the default language instead.}%
\else
\language=\csname l@#1\endcsname
\fi
#2}}
\providecommand{\BIBdecl}{\relax}
\BIBdecl

\bibitem{zhang2024model}
H.~Zhang, M.~Chen, A.~Vahid, and H.~Sun, ``Model-based deep learning for rate split multiple access in vehicular communications,'' \emph{arXiv preprint arXiv:2405.01515}, 2024.

\bibitem{zhang2024transformer}
S.~Zhang, S.~Zhang, Y.~Mao, L.~K. Yeung, B.~Clerckx, and T.~Q. Quek, ``Transformer-based channel prediction for rate-splitting multiple access-enabled vehicle-to-everything communication,'' \emph{IEEE Transactions on Wireless Communications}, 2024.

\bibitem{zhuang2018large}
B.~Zhuang, D.~Guo, E.~Wei, and M.~L. Honig, ``Large-scale spectrum allocation for cellular networks via sparse optimization,'' \emph{IEEE Transactions on Signal Processing}, vol.~66, no.~20, pp. 5470--5483, 2018.

\bibitem{mathur2021survey}
H.~Mathur and T.~Deepa, ``A survey on advanced multiple access techniques for 5g and beyond wireless communications,'' \emph{Wireless Personal Communications}, vol. 118, no.~2, pp. 1775--1792, 2021.

\bibitem{alizadeh2023power}
M.~Alizadeh and H.~Tabassum, ``Power control with qos guarantees: A differentiable projection-based unsupervised learning framework,'' \emph{IEEE Transactions on Communications}, vol.~71, no.~8, pp. 4605--4619, 2023.

\bibitem{mao2018rate}
Y.~Mao, B.~Clerckx, and V.~O. Li, ``Rate-splitting multiple access for downlink communication systems: Bridging, generalizing, and outperforming sdma and noma,'' \emph{EURASIP journal on wireless communications and networking}, vol. 2018, pp. 1--54, 2018.

\bibitem{yang2021optimization}
Z.~Yang, M.~Chen, W.~Saad, and M.~Shikh-Bahaei, ``Optimization of rate allocation and power control for rate splitting multiple access (rsma),'' \emph{IEEE Transactions on Communications}, vol.~69, no.~9, pp. 5988--6002, 2021.

\bibitem{clerckx2023primer}
B.~Clerckx, Y.~Mao, E.~A. Jorswieck, J.~Yuan, D.~J. Love, E.~Erkip, and D.~Niyato, ``A primer on rate-splitting multiple access: Tutorial, myths, and frequently asked questions,'' \emph{IEEE Journal on Selected Areas in Communications}, 2023.

\bibitem{clerckx2016rate}
B.~Clerckx, H.~Joudeh, C.~Hao, M.~Dai, and B.~Rassouli, ``Rate splitting for mimo wireless networks: A promising phy-layer strategy for lte evolution,'' \emph{IEEE Communications Magazine}, vol.~54, no.~5, pp. 98--105, 2016.

\bibitem{clerckx2024multiple}
B.~Clerckx, Y.~Mao, Z.~Yang, M.~Chen, A.~Alkhateeb, L.~Liu, M.~Qiu, J.~Yuan, V.~W. Wong, and J.~Montojo, ``Multiple access techniques for intelligent and multi-functional 6g: Tutorial, survey, and outlook,'' \emph{arXiv preprint arXiv:2401.01433}, 2024.

\bibitem{xiao2023joint}
M.~Xiao, H.~Cui, Z.~Zhao, X.~Cao, and D.~O. Wu, ``Joint 3d deployment and beamforming for rsma-enabled uav base station with geographic information,'' \emph{IEEE Transactions on Wireless Communications}, 2023.

\bibitem{dizdar2023rsma}
O.~Dizdar, A.~Sattarzadeh, Y.~X. Yap, and S.~Wang, ``Rsma for overloaded mimo networks: Low-complexity design for max-min fairness,'' \emph{IEEE Transactions on Wireless Communications}, 2023.

\bibitem{gong2023edge}
T.~Gong, L.~Zhu, F.~R. Yu, and T.~Tang, ``Edge intelligence in intelligent transportation systems: A survey,'' \emph{IEEE Transactions on Intelligent Transportation Systems}, vol.~24, no.~9, pp. 8919--8944, 2023.

\bibitem{zhou2022deep}
T.~Zhou, H.~Zhang, B.~Ai, C.~Xue, and L.~Liu, ``Deep-learning-based spatial--temporal channel prediction for smart high-speed railway communication networks,'' \emph{IEEE Transactions on Wireless Communications}, vol.~21, no.~7, pp. 5333--5345, 2022.

\bibitem{wang2020deep}
Y.~Wang, J.~Wang, W.~Zhang, J.~Yang, and G.~Gui, ``Deep learning-based cooperative automatic modulation classification method for mimo systems,'' \emph{Ieee transactions on vehicular technology}, vol.~69, no.~4, pp. 4575--4579, 2020.

\bibitem{zhang2022learning}
Z.~Zhang, T.~Jiang, and W.~Yu, ``Learning based user scheduling in reconfigurable intelligent surface assisted multiuser downlink,'' \emph{IEEE Journal of Selected Topics in Signal Processing}, vol.~16, no.~5, pp. 1026--1039, 2022.

\bibitem{li2023graph}
X.~Li, M.~Chen, Y.~Liu, Z.~Zhang, D.~Liu, and S.~Mao, ``Graph neural networks for joint communication and sensing optimization in vehicular networks,'' \emph{IEEE Journal on Selected Areas in Communications}, 2023.

\bibitem{liu2021towards}
J.~Liu, Z.~Shen, Y.~He, X.~Zhang, R.~Xu, H.~Yu, and P.~Cui, ``Towards out-of-distribution generalization: A survey,'' \emph{arXiv preprint arXiv:2108.13624}, 2021.

\bibitem{robey2020model}
A.~Robey, H.~Hassani, and G.~J. Pappas, ``Model-based robust deep learning: Generalizing to natural, out-of-distribution data,'' \emph{arXiv preprint arXiv:2005.10247}, 2020.

\bibitem{shlezinger2023model}
N.~Shlezinger, J.~Whang, Y.~C. Eldar, and A.~G. Dimakis, ``Model-based deep learning,'' \emph{Proceedings of the IEEE}, 2023.

\bibitem{samuel2019learning}
N.~Samuel, T.~Diskin, and A.~Wiesel, ``Learning to detect,'' \emph{IEEE Transactions on Signal Processing}, vol.~67, no.~10, pp. 2554--2564, 2019.

\bibitem{nguyen2020deep}
N.~T. Nguyen and K.~Lee, ``Deep learning-aided tabu search detection for large mimo systems,'' \emph{IEEE Transactions on Wireless Communications}, vol.~19, no.~6, pp. 4262--4275, 2020.

\bibitem{zhang2023joint}
J.~Zhang, C.~Masouros, and L.~Hanzo, ``Joint precoding and csi dimensionality reduction: An efficient deep unfolding approach,'' \emph{IEEE Transactions on Wireless Communications}, 2023.

\bibitem{xia2023deep}
W.~Xia, Y.~Jiang, B.~Zhao, H.~Zhao, and H.~Zhu, ``Deep unfolded fractional programming based beamforming in ris-aided miso systems,'' \emph{IEEE Wireless Communications Letters}, 2023.

\bibitem{schynol2023coordinated}
L.~Schynol and M.~Pesavento, ``Coordinated sum-rate maximization in multicell mu-mimo with deep unrolling,'' \emph{IEEE Journal on Selected Areas in Communications}, vol.~41, no.~4, pp. 1120--1134, 2023.

\bibitem{zhang2022deep}
Q.~Zhang, L.~Zhu, S.~Jiang, and X.~Tang, ``Deep unfolding for cooperative rate splitting multiple access in hybrid satellite terrestrial networks,'' \emph{China Communications}, vol.~19, no.~7, pp. 100--109, 2022.

\bibitem{NEURIPS2023_47547ee8}
\BIBentryALTinterwordspacing
F.~Zhong, K.~Fogarty, P.~Hanji, T.~Wu, A.~Sztrajman, A.~Spielberg, A.~Tagliasacchi, P.~Bosilj, and C.~Oztireli, ``Neural fields with hard constraints of arbitrary differential order,'' in \emph{Advances in Neural Information Processing Systems}, A.~Oh, T.~Naumann, A.~Globerson, K.~Saenko, M.~Hardt, and S.~Levine, Eds., vol.~36.\hskip 1em plus 0.5em minus 0.4em\relax Curran Associates, Inc., 2023, pp. 22\,870--22\,895. [Online]. Available: \url{https://proceedings.neurips.cc/paper_files/paper/2023/file/47547ee84e3fbbcbbbbad7c1fd9a973b-Paper-Conference.pdf}
\BIBentrySTDinterwordspacing

\bibitem{cristian2023end}
R.~Cristian, P.~Harsha, G.~Perakis, B.~L. Quanz, and I.~Spantidakis, ``End-to-end learning for optimization via constraint-enforcing approximators,'' in \emph{Proceedings of the AAAI Conference on Artificial Intelligence}, vol.~37, no.~6, 2023, pp. 7253--7260.

\bibitem{gao2022online}
J.~Gao, C.~Zhong, G.~Y. Li, and Z.~Zhang, ``Online deep neural network for optimization in wireless communications,'' \emph{IEEE Wireless Communications Letters}, vol.~11, no.~5, pp. 933--937, 2022.

\bibitem{zhang2020deep}
J.~Zhang, W.~Xia, M.~You, G.~Zheng, S.~Lambotharan, and K.-K. Wong, ``Deep learning enabled optimization of downlink beamforming under per-antenna power constraints: Algorithms and experimental demonstration,'' \emph{IEEE Transactions on Wireless Communications}, vol.~19, no.~6, pp. 3738--3752, 2020.

\bibitem{huang2023deep}
J.~Huang, Y.~Yang, J.~Lee, D.~He, and Y.~Li, ``Deep reinforcement learning based resource allocation for rsma in leo satellite-terrestrial networks,'' \emph{IEEE Transactions on Communications}, 2023.

\bibitem{ye2019deep}
H.~Ye, G.~Y. Li, and B.-H.~F. Juang, ``Deep reinforcement learning based resource allocation for v2v communications,'' \emph{IEEE Transactions on Vehicular Technology}, vol.~68, no.~4, pp. 3163--3173, 2019.

\bibitem{wang2023graph}
Z.~Wang, Y.~Zhou, Y.~Zou, Q.~An, Y.~Shi, and M.~Bennis, ``A graph neural network learning approach to optimize ris-assisted federated learning,'' \emph{IEEE Transactions on Wireless Communications}, 2023.

\bibitem{xu2023distributed}
X.~Xu, Y.~Liu, Q.~Chen, X.~Mu, and Z.~Ding, ``Distributed auto-learning gnn for multi-cell cluster-free noma communications,'' \emph{IEEE Journal on Selected Areas in Communications}, vol.~41, no.~4, pp. 1243--1258, 2023.

\bibitem{zhang2023gnn}
H.~Zhang, X.~Ma, X.~Liu, L.~Li, and K.~Sun, ``Gnn-based power allocation and user association in digital twin network for the terahertz band,'' \emph{IEEE Journal on Selected Areas in Communications}, 2023.

\bibitem{shi2011iteratively}
Q.~Shi, M.~Razaviyayn, Z.-Q. Luo, and C.~He, ``An iteratively weighted mmse approach to distributed sum-utility maximization for a mimo interfering broadcast channel,'' \emph{IEEE Transactions on Signal Processing}, vol.~59, no.~9, pp. 4331--4340, 2011.

\bibitem{liang2019towards}
F.~Liang, C.~Shen, W.~Yu, and F.~Wu, ``Towards optimal power control via ensembling deep neural networks,'' \emph{IEEE Transactions on Communications}, vol.~68, no.~3, pp. 1760--1776, 2019.

\bibitem{eisen2020optimal}
M.~Eisen and A.~Ribeiro, ``Optimal wireless resource allocation with random edge graph neural networks,'' \emph{ieee transactions on signal processing}, vol.~68, pp. 2977--2991, 2020.

\bibitem{naderializadeh2020wireless}
N.~Naderializadeh, M.~Eisen, and A.~Ribeiro, ``Wireless power control via counterfactual optimization of graph neural networks,'' in \emph{2020 IEEE 21st international workshop on signal processing advances in wireless communications (SPAWC)}.\hskip 1em plus 0.5em minus 0.4em\relax IEEE, 2020, pp. 1--5.

\bibitem{li2021multicell}
Y.~Li, S.~Han, and C.~Yang, ``Multicell power control under rate constraints with deep learning,'' \emph{IEEE Transactions on Wireless Communications}, vol.~20, no.~12, pp. 7813--7825, 2021.

\bibitem{shen2018fractional}
K.~Shen and W.~Yu, ``Fractional programming for communication systems—part i: Power control and beamforming,'' \emph{IEEE Transactions on Signal Processing}, vol.~66, no.~10, pp. 2616--2630, 2018.

\bibitem{grant2014cvx}
M.~Grant and S.~Boyd, ``Cvx: Matlab software for disciplined convex programming, version 2.1,'' 2014.

\bibitem{szegedy2015going}
C.~Szegedy, W.~Liu, Y.~Jia, P.~Sermanet, S.~Reed, D.~Anguelov, D.~Erhan, V.~Vanhoucke, and A.~Rabinovich, ``Going deeper with convolutions,'' in \emph{Proceedings of the IEEE conference on computer vision and pattern recognition}, 2015, pp. 1--9.

\bibitem{chen2019deep}
M.~Chen, J.~Chen, X.~Chen, S.~Zhang, and S.~Xu, ``A deep learning based resource allocation scheme in vehicular communication systems,'' in \emph{2019 IEEE Wireless Communications and Networking Conference (WCNC)}.\hskip 1em plus 0.5em minus 0.4em\relax IEEE, 2019, pp. 1--6.

\bibitem{lei2019learning}
L.~Lei, Y.~Yuan, T.~X. Vu, S.~Chatzinotas, and B.~Ottersten, ``Learning-based resource allocation: Efficient content delivery enabled by convolutional neural network,'' in \emph{2019 IEEE 20th International Workshop on Signal Processing Advances in Wireless Communications (SPAWC)}.\hskip 1em plus 0.5em minus 0.4em\relax IEEE, 2019, pp. 1--5.

\end{thebibliography}

\end{document}